\documentclass[acmsmall]{acmart}

\usepackage{acro}
\DeclareAcronym{soc}{
  short=SoC,
  long=System-on-Chip,
}
\DeclareAcronym{socks}{
  short=\textit{SoCks},
  long=System-on-Chip blocks,
}
\DeclareAcronym{os}{
  short=OS,
  long=Operating System,
}
\DeclareAcronym{fpga}{
  short=FPGA,
  long=Field-Programmable Gate Array,
}
\DeclareAcronym{cicd}{
  short=CI/CD,
  long=Continuous Integration and Continuous Delivery,
}
\DeclareAcronym{adc}{
  short=ADC,
  long=Analog-to-Digital Converter,
}
\DeclareAcronym{dac}{
  short=DAC,
  long=Digital-to-Analog Converter,
}
\DeclareAcronym{rfsoc}{
  short=RFSoC,
  long=Radio-Frequency System-on-Chip,
}
\DeclareAcronym{mpsoc}{
  short=MPSoC,
  long=Multiprocessor System-on-Chip,
}
\DeclareAcronym{cpu}{
  short=CPU,
  long=Central Processing Unit,
}
\DeclareAcronym{xsa}{
  short=XSA,
  long=Xilinx Support Archive,
}
\DeclareAcronym{sh}{
  short=sh,
  long=Bourne Shell,
}
\DeclareAcronym{bash}{
  short=Bash,
  long=Bourne Again Shell,
}
\DeclareAcronym{posix}{
  short=POSIX,
  long=Portable Operating System Interface,
}
\DeclareAcronym{gui}{
  short=GUI,
  long=Graphical User Interface,
}
\DeclareAcronym{cli}{
  short=CLI,
  long=Command-Line Interface,
}
\DeclareAcronym{ide}{
  short=IDE,
  long=Integrated Development Environment,
}
\DeclareAcronym{url}{
  short=URL,
  long=Uniform Resource Locator,
}
\DeclareAcronym{csv}{
  short=CSV,
  long=Comma-Separated Values,
}
\DeclareAcronym{nvme}{
  short=NVMe,
  long=Non-Volatile Memory Express,
}
\DeclareAcronym{ssd}{
  short=SSD,
  long=Solid-State Drive,
}
\DeclareAcronym{rhel}{
  short=RHEL,
  long=Red Hat Enterprise Linux,
}
\DeclareAcronym{sata}{
  short=SATA,
  long=Serial AT Attachment,
}
\DeclareAcronym{ai}{
  short=AI,
  long=Artificial Intelligence,
}
\DeclareAcronym{ipbb}{
  short=IPBB,
  long=IPbus builder,
}
\DeclareAcronym{us+}{
  short=US+,
  long=UltraScale+,
}
\DeclareAcronym{sdr}{
  short=SDR,
  long=Software-Defined Radio,
}
\DeclareAcronym{gpu}{
  short=GPU,
  long=Graphics Processing Unit,
}
\DeclareAcronym{qubit}{
  short=qubit,
  long=quantum bit,
}
\DeclareAcronym{cms}{
  short=CMS,
  long=Compact Muon Solenoid,
}
\DeclareAcronym{hllhc}{
  short=HL-LHC,
  long=High-Luminosity Large Hadron Collider,
}
\DeclareAcronym{hep}{
  short=HEP,
  long=High Energy Physics,
}
\DeclareAcronym{ram}{
  short=RAM,
  long=Random-Access Memory,
}

\usepackage{minted}

\AtBeginDocument{%
  }

\begin{document}

\title{SoCks — Simplifying Firmware and Software Integration for Heterogeneous SoCs}

\author{Marvin Fuchs}
\email{marvin.fuchs@kit.edu}
\orcid{0000-0002-4146-5846}
\author{Lukas Scheller}
\orcid{0009-0003-9156-7781}
\author{Timo Muscheid}
\orcid{0000-0002-1108-7784}
\author{Oliver Sander}
\orcid{0000-0002-0959-4744}
\author{Luis E. Ardila-Perez}
\orcid{0000-0002-7485-8267}
\affiliation{%
  \institution{Institute for Data Processing and Electronics, Karlsruhe Institute of Technology}
  \city{Karlsruhe}
  \country{Germany}
}

\renewcommand{\shortauthors}{Fuchs et al.}

\begin{abstract}
Modern heterogeneous \acf{soc} devices integrate advanced components into a single package, offering powerful capabilities while also introducing significant complexity. To manage these sophisticated devices, firmware and software developers need powerful development tools. However, as these tools become increasingly complex, they often lack adequate support, resulting in a steep learning curve and challenging troubleshooting. To address this, this work introduces \acf{socks}, a flexible and expandable build framework that reduces complexity by partitioning the \acs{soc} image into high-level units called blocks. \acs{socks} builds each firmware and software block in an encapsulated way, independently from other components of the image, thereby reducing dependencies to a minimum. While some information exchange between the blocks is unavoidable to ensure seamless runtime integration, this interaction is standardized via interfaces. A small number of dependencies and well-defined interfaces simplify the reuse of existing block implementations and facilitate seamless substitution between versions---for instance, when choosing root file systems for the embedded Linux operating system. Additionally, this approach facilitates the establishment of a decentralized and partially automated development flow through \ac{cicd}. Measurement results demonstrate that \acs{socks} can build a complete \acs{soc} image up to three times faster than established tools.
\end{abstract}



\received{24 September 2025}

\maketitle

\section{Introduction}
Advances in modern microelectronics manufacturing processes with increased integration density have enabled powerful heterogeneous \ac{soc} devices, combining various components such as processors, \acp{gpu}, \ac{ai} accelerators, \ac{fpga} fabric, and high-speed interfaces into a single package. Tight integration within these chips enables internal communication with low latency and high data throughput while keeping the system's energy consumption low. Since the flexibility of such single-chip systems is generally limited by their immutable composition, there are a variety of designs for different use cases. One area of application is mobile computing devices such as smartphones, tablets, and laptop computers. These devices leverage high integration density to improve performance per watt efficiency in a compact packaging format at low cost. Beyond mobile computing, consumer goods like cars, drones, and smart TVs are increasingly using \acp{soc}, specifically designed for these fields, to reduce cost and incorporate the latest technologies, such as advanced online data processing and \ac{ai} \cite{NvidiaThor, QualcommQRB5165, MediaTekPentonic2000}.

All of the application areas discussed so far share two key commonalities: their devices have very similar functional requirements and are produced at scale. This makes developing specialized \acp{soc} for these fields both feasible and lucrative. To harness the advantages of \ac{soc} devices in a wider range of applications with greater flexibility, manufacturers have begun to integrate programmable logic using \ac{fpga} technology. The result allows developers to add application-specific hardware accelerators, soft processors, and interfaces as needed. Compared to fixed-logic \acp{soc}, the additional flexibility comes at the cost of lower energy efficiency, reduced integration density, and higher prices. In contrast, when comparing \ac{fpga}-assisted \acp{soc} to standalone \acp{fpga}, the trade-offs are reversed. Hybrid \ac{soc}-\ac{fpga} devices typically offer better energy efficiency and higher integration density at similar costs. This makes these heterogeneous devices an attractive choice for areas where powerful \acp{fpga} have already established themselves.

One such field is the development of highly specialized electronics for fundamental physics experiments. In \ac{hep}, \acp{fpga} implement constantly evolving algorithms for real-time data processing \cite{Phase2_Upgrade_Tracker}. Regular algorithm updates are crucial, for example, to improve particle track reconstruction performance, which is essential for obtaining the highest-quality measurement data. Serenity-S---a high-throughput processing card developed for such applications in the Phase-2 Upgrade of the \ac{cms} detector at the \ac{hllhc}---leverages a Zynq \ac{us+} \ac{mpsoc} for advanced hardware management and monitoring along with a Virtex Ultrascale+ VU13P \ac{fpga} for online data processing \cite{Mehner_2024}. A dedicated \ac{fpga} is necessary here, as the \ac{fpga} resources within the \ac{mpsoc} are insufficient. Nevertheless, an \ac{fpga}-assisted \acp{soc} is still used because its heterogeneous architecture enables seamless integration into computer networks, robust connectivity to onboard devices, and support for custom interfaces through the internal \ac{fpga} fabric. In future developments, such resource expansions with dedicated \acp{fpga} will be less necessary, as the newer Versal family of AMD \acp{soc} offers significantly more resources than the Zynq \ac{us+}. A recent survey conducted at CERN shows that the use of \acp{soc} for such and similar applications is common practice in the \ac{hep} community \cite{CERNSoCStudy}. Furthermore, there are also state-of-the-art instruments that fit entirely into a single heterogeneous high-performance \ac{soc}. One example is the QiController---a control platform for superconducting \acp{qubit}---based on a single \ac{rfsoc} device from AMD. The QiController utilizes a Linux \ac{os} running on the hardened processors for system management and user interface, while computationally intensive, high-bandwidth, and real-time tasks are handled by custom logic in the \ac{fpga} fabric. Integrated \acp{dac} and \acp{adc} enable direct interfacing with the analog signals required to readout and control the superconducting \ac{qubit} devices.

In addition to their technical capabilities, \acp{soc} are also attractive because they simplify development. While integrating a system from discrete components is a major challenge that requires solving power management, clocking, and interface issues at a very low level, most of these challenges are already solved by the manufacturer in an \ac{soc}. Furthermore, the software-defined behavior of these devices allows for significant changes, even late in development or during operation. This flexibility makes it easier to adapt an already existing system to new requirements, or share components between multiple devices.

This work is primarily motivated by the increasing use of heterogeneous high-performance \ac{soc} devices in fundamental physics research. Typical examples are the Zynq, Zynq \ac{us+}, and Versal families of devices from AMD. The images required to operate such \acp{soc} are typically created with the development tools provided by the manufacturer, commonly based on open-source build frameworks for embedded Linux environments such as Yocto and Buildroot \cite{BuildrootWeb, YoctoWeb}. For example, AMD offers the Yocto-based Petalinux Tools for development but recommends using pure Yocto for production \cite{PetaLinux2Prod}. These frameworks are designed to build a fully custom \ac{os} that can be trimmed to the bare minimum to run on systems with very limited resources. However, high-performance \acp{soc} typically do not have such limitations and are powerful enough to run regular Linux distributions, similar to a Raspberry Pi. Utilizing a regular Linux distribution brings a number of advantages, such as public repositories and regular updates, which can ease both the development of the image and the operation of the system significantly.

In addition to the central \ac{os}, bootable images for modern heterogeneous \ac{soc} devices include various application- and architecture-specific binary files. Examples range from low-level firmware, bootloaders, and userspace applications to \ac{fpga} firmware. Building these files from source is in most cases not trivial and requires powerful development tools like compilers, build systems, and \ac{fpga} synthesis software. Given the complexity of these tools, it is not feasible to implement all their functions in a single, comprehensive tool. Instead, both established build frameworks, Buildroot and Yocto, act as a superordinate structure that uses underlying tools specialized in building individual components. This work adopts this proven concept but combines it with a new approach by grouping firmware and software components according to their build-time and runtime relationships, thereby minimizing dependencies between components.

In this article, we introduce \ac{socks}, a modular build system for bootable images designed for high-performance \acp{soc} running an embedded Linux \ac{os}. Particular attention is paid to the implementation of the aforementioned concept, which breaks down the \ac{soc} image into groups of components, the so-called blocks. Introducing these blocks enables modularization of \ac{soc} images at a new level of abstraction. We show how this additional level of abstraction helps simplify the build process of full \ac{soc} images while making it faster and less demanding for the build PC in terms of \ac{cpu} power, memory, and disk space compared to common build frameworks. The improvement of the build process is underlined by measurement results. Methods for building the blocks and their integration into a complete image are discussed. Finally, practices facilitated by \ac{socks} for efficient and distributed development, such as the reuse of existing components or automated building and testing with \ac{cicd}, are presented.

\section{Related Work}

By far the most widely used frameworks to build images for \ac{fpga}-assisted \ac{soc} devices are Buildroot and Yocto. One of the main reasons for this is the official support from market-leading manufacturers for both build systems. Both AMD and Microchip provide support for building images using Buildroot and Yocto, while Altera only offers support for Yocto \cite{AMDBuildroot, PolarfireSW, AlteraFPGAUserGuide}.

Both frameworks are designed to build custom Linux-based systems, primarily for embedded devices. In contrast to general-purpose machines such as laptops and servers, embedded systems often use a custom Linux \ac{os} tailored to their specific needs. This feature is due to the often severely limited resources of such devices, which make it essential to strip the \ac{os} to the bare minimum. Another distinguishing characteristic of embedded systems is their typically application-specific hardware architecture. A standardized hardware description that can be evaluated at runtime, such as a devicetree, is therefore critical for embedded devices. This is especially true for \acp{soc} that include \ac{fpga} fabric, allowing them to modify their hardware composition at runtime in specific ways. Finally, embedded systems often operate unseen by users, so they must function reliably and autonomously, without user interaction or regular updates, for months or even years. This situation leads to significant challenges for the security and stability of such systems.

\begin{figure}[b]
  \centering
  \includegraphics[width=\linewidth]{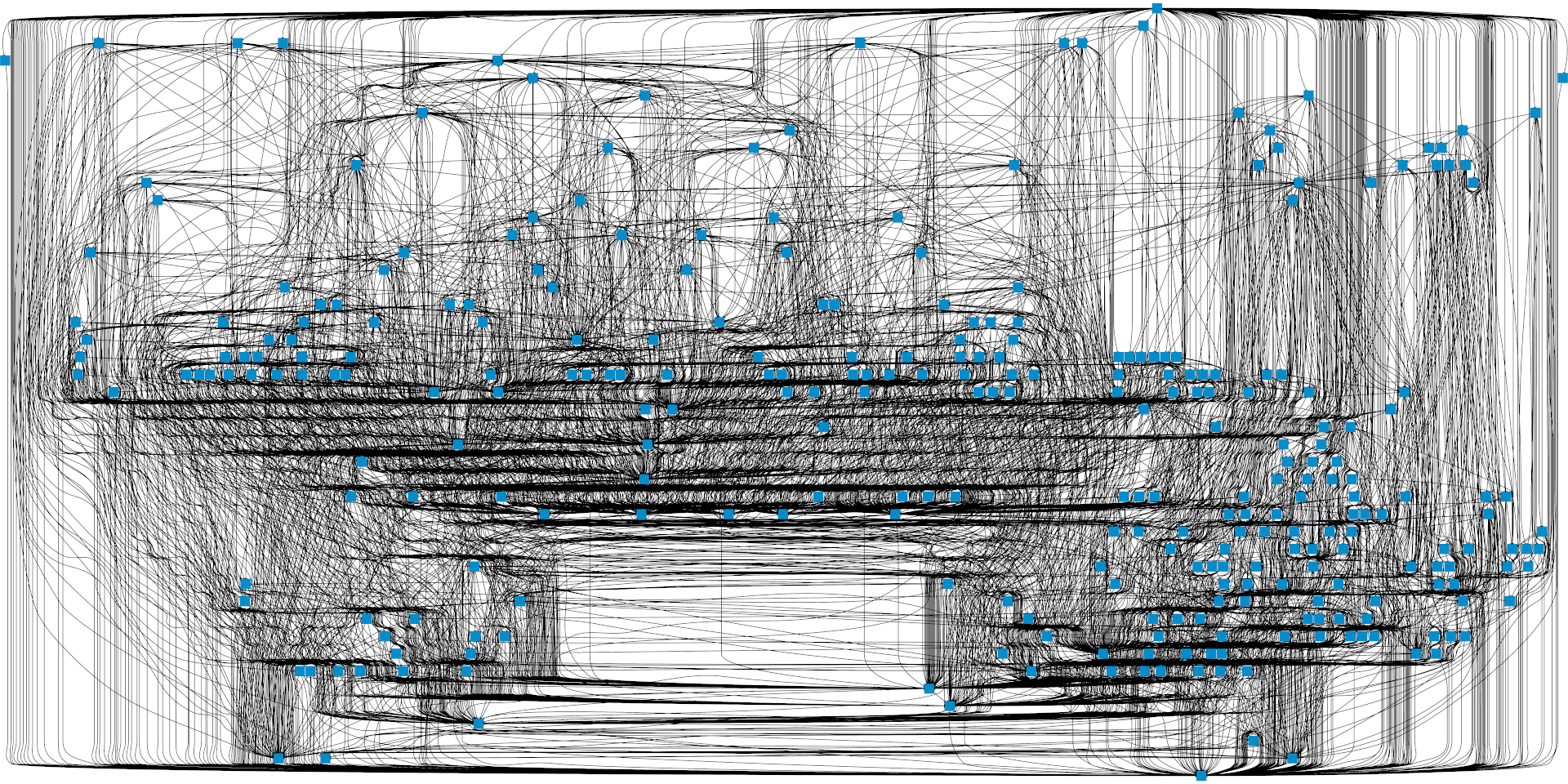}
  \caption{Dependency graph of a Yocto project targeting an AMD Zynq \ac{us+} \ac{mpsoc} \cite{fuchs_2025_17131066}. Each blue rectangle in the graph represents one BitBake recipe. The edges between the recipes represent the dependencies.}
  \label{fig:yocto_image_dependencies}
\end{figure}

Addressing all of these requirements while maintaining full flexibility and providing support for the multitude of \ac{cpu} instruction sets that are frequently used in \ac{soc} devices imposes significant challenges on a development framework. To meet these challenges, both Yocto and Buildroot take a fundamentally different approach than traditional Linux distributions. They do not rely on publicly available repositories to distribute binary packages that can extend the functionality of the \ac{os} at runtime \cite{YoctoPackageDocu, BuildrootPackageDocu}. Runtime package management is often omitted entirely, resulting in a so-called static Linux system. Instead, these frameworks build almost everything---including the required toolchains---from source. Together with the abstract and open design of Yocto and Buildroot, this allows for extensive customization of individual components, their build process, and their composition into an image, enabling support for a wide range of \ac{soc} architectures and use cases. However, this flexibility comes at a cost: it leads to a multitude of complex configuration files, deeply nested project structures, and opaque dependencies. Fig.~\ref{fig:yocto_image_dependencies} gives an impression of this. It shows all BitBake recipes used in a Yocto project for an AMD Zynq \ac{us+} device and their dependencies. Each recipe (.bb file) defines in detail how a component is built, its dependencies, and how it is integrated into the image. In addition, append files (.bbappend)---which are not shown in the figure---can modify and extend recipes. The example in Fig.~\ref{fig:yocto_image_dependencies} uses only a fraction of the more than 11,000 recipes and more than 2,000 append files available in the project. This extensive configurability places significant responsibility on the user, potentially blurring the line between the user project and the development tool.

For high-performance \acp{soc}, this level of effort is usually not required, because they are powerful enough to run a regular Linux distribution, similar to Raspberry Pi \ac{os}. Using a regular distribution with a package manager, public repositories, and regular updates can significantly simplify the development and operation of an embedded \ac{os}. However, neither Buildroot nor Yocto supports such an approach.

Furthermore, a Yocto or Buildroot project for an \ac{fpga}-assisted \ac{soc}, such as an AMD Zynq \ac{us+} device, is generally not completely self-contained. Since both frameworks focus on building an embedded Linux system, they are not intended for low-level \ac{soc} configuration. With AMD Zynq, Zynq \ac{us+}, and Versal devices, such configurations are typically made in the Vivado design suite. Vivado is primarily a synthesis tool for hardware description languages, but it provides a multitude of other functions beyond that. These include low-level configurations of the \acp{soc}, such as clock frequencies, utilized interfaces, and power management. The configuration, together with a bitfile for the \ac{fpga} fabric, can be exported as an \ac{xsa} file. This archive can then be imported and further processed by Yocto and Buildroot, resulting in a two-stage development process that typically has to be carried out manually. By creating a custom Yocto recipe or Buildroot package and utilizing Vivado's Tcl API, it is possible to merge the two stages. However, due to the typically long build times of Vivado projects---which can take several hours---developers generally prefer to stay in control and decide for themselves when and how to build a Vivado project.

Beyond these two widely used frameworks, there is little active research on building frameworks for heterogeneous high-performance \acp{soc}. In the field of \ac{soc} devices, research mainly focuses on the development of these devices themselves, their hardware architecture, and layout \cite{Kermarrec_2020, Shalan_2020, Mantovani_2020, Amid_2020, Kangas_2006}. An exception is FireMarshal, a development tool capable of building and simulating bootable images for \ac{soc} devices, inspired by the computer architecture community \cite{Pemberton_2021}. However, this tool is specifically designed to support the development, benchmarking, and comparison of new architectures and is therefore not optimized for production-ready images. Researchers who need to build production-ready images frequently rely on custom scripted workflows that are tailored to the specific project \cite{DIOTImageFramework}. To fully automate such workflows, there is also activity in developing wrappers that integrate existing development tools, such as AMD's Vivado \cite{Hog_2021, IPbusBuilder, Karcher_2022}. Such approaches are sometimes the only option, because the low-level implementation details of market-leading high-performance \acp{soc} are proprietary, and as a result, the development of a completely independent development tool would only be possible through extensive reverse engineering. Nevertheless, the development of such wrappers can be very fruitful, as they can significantly improve the usability of existing tools while keeping the development effort low.


\section{Concept}
\label{sec:concept}

To avoid the overwhelming complexity involved in creating an \ac{soc} image with the existing tools, \ac{socks} follows a different, less flexible approach that focuses on specific \ac{soc} architectures and enforces a strict separation between user project and development tool. Furthermore, the tool is lightweight by design, and the behavior of \ac{socks} is always transparent to the user. However, the primary strategy for reducing complexity in \ac{socks} is to partition \ac{soc} images into a reasonable number of manageable units that can be treated mostly independently. The number of partitions is critical: too few large partitions do not reduce the complexity of the individual units significantly, whereas too many small partitions result in excessive dependencies, leading to additional complexity. The optimal number depends on the architecture of the \ac{soc} and the firmware and software infrastructure that is used on it. It should therefore not be defined uniformly for all \ac{soc} architectures by the \ac{socks} framework. Finally, partitioning should be done at an intuitive abstraction layer, ensuring developers can manage the components naturally.

\begin{figure}[t]
  \centering
  \includegraphics[width=\linewidth]{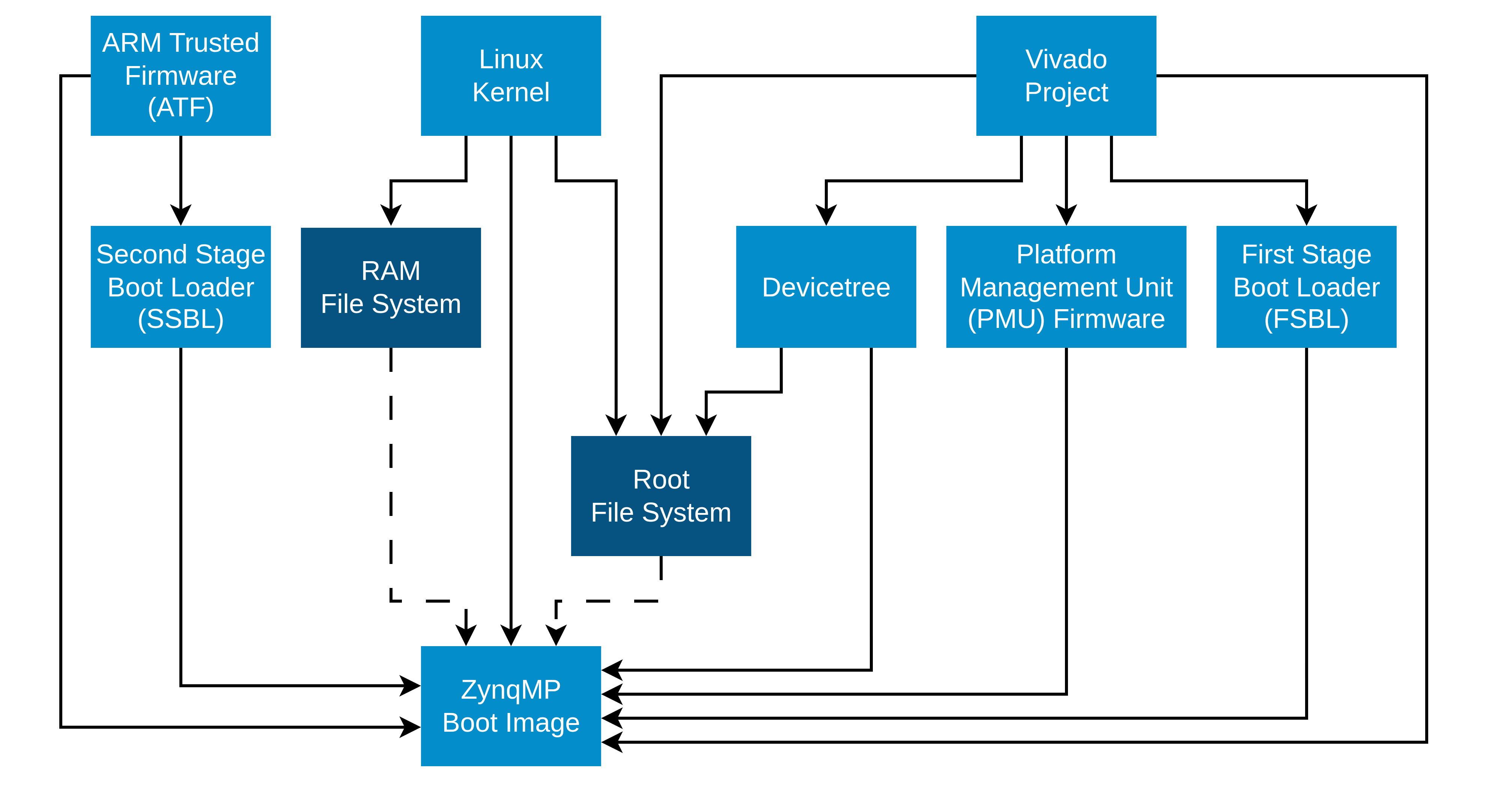}
  \caption{Data flow graph of a \ac{socks} project targeting an AMD Zynq \ac{us+} \ac{mpsoc}. This example shows the partitioning of a complete image at the abstraction level of \ac{socks}. ``RAM File System'' and ``Root File System'' are optional blocks and therefore dark blue. An \ac{soc} image utilizing an embedded Linux \ac{os} can have either one of the two file systems or both. The latter approach corresponds to a Linux \ac{os} with an initramfs.}
  \label{fig:partitioned_image}
\end{figure}

Partitioning the image into defined segments with a limited number of dependencies reduces complexity and enables a modular approach. Modularization makes it easier to divide development tasks among several developers and facilitates the reuse of existing components, which can further accelerate the development of \ac{soc} images. To take full advantage of the modular approach, the partitioning of the image is fixed for a given \ac{soc} architecture, and different modules for the same partition must be interchangeable. To enable this, it is required to define standardized interfaces for the modules. In \ac{socks}, such a module with standardized interfaces is called an ``\ac{soc} block''. The interfaces enable unidirectional data transfer between two blocks, allowing multiple blocks to be combined into a complete \ac{soc} image. Depending on the software infrastructure used on the \ac{soc}, it can be sensible to make some blocks optional. For instance, it is possible to employ a Linux \ac{os} with a non-persistent \ac{ram} file system, with a persistent root file system, or with a combination of both. Fig.~\ref{fig:partitioned_image} shows the partitioned image for an AMD Zynq \ac{us+} \ac{mpsoc}. For most of the blocks in the figure, their name already indicates which part of the \ac{soc} image they represent. One exception is the ``Vivado Project'' block, which includes the \ac{soc}'s low-level hardware configuration and the configuration file for the embedded \ac{fpga} fabric. The figure indicates that there are multiple layers of dependencies between the blocks and that the ``Boot Image'' block combines data from all other blocks to form the bootable image. By comparing Fig.~\ref{fig:yocto_image_dependencies} and Fig.~\ref{fig:partitioned_image}, the reduced complexity of a \ac{socks} project becomes apparent.

\begin{figure}[t]
  \centering
  \includegraphics[width=\linewidth]{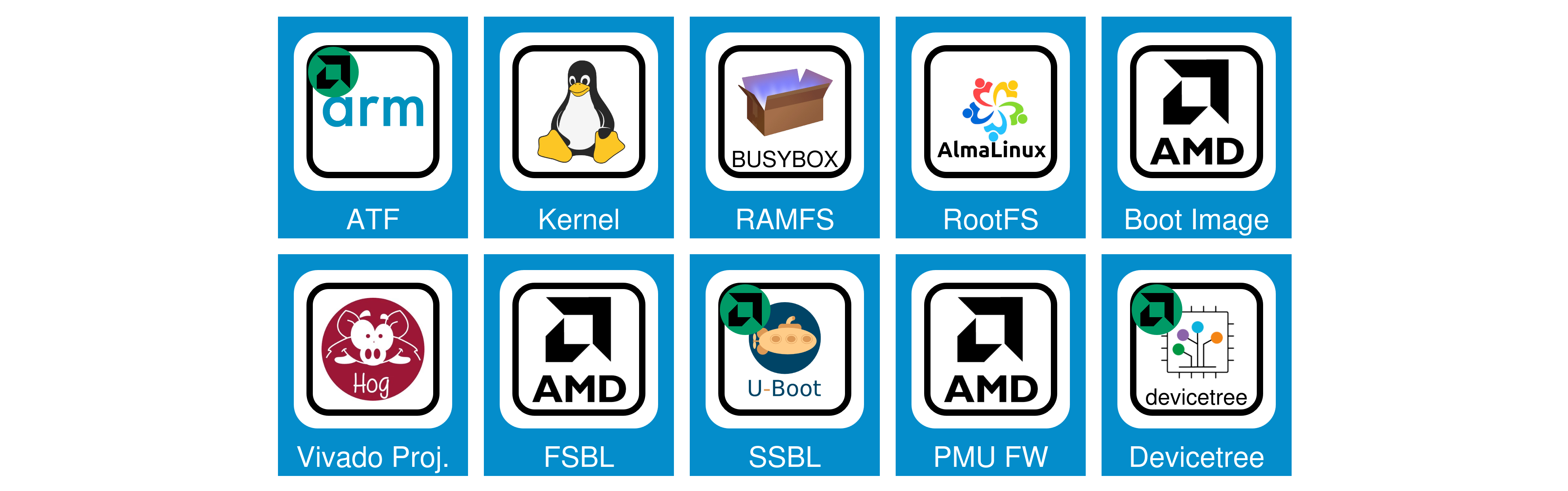}
  \caption{Blocks of an AMD Zynq \ac{us+} \ac{mpsoc} image with associated builders. Tiles with an AMD icon in the top left corner indicate that the builder is optimized for source code that has been adapted by the manufacturer for the target architecture.}
  \label{fig:partitioned_blocks}
\end{figure}

In most cases, there are several ways in which the content of a block can be implemented and built. For instance, there are multiple versions of the Linux kernel, a variety of root file system flavors, and frameworks that store a Vivado project in a Git-compatible format and enable fully automated building, like Hog, \ac{ipbb}, and logicc \cite{Hog_2021, IPbusBuilder, Karcher_2022}. In \ac{socks}, one builder must be assigned to each block, representing a specific way in which the block can be created. A complete \ac{soc} image with builders selected for every block is depicted in Fig.~\ref{fig:partitioned_blocks}. In the case of the root file system, a Debian builder is also conceivable in addition to the AlmaLinux builder shown. Switching between these two builders---equivalent to switching between two Linux distributions---is straightforward due to the standardized interfaces. This addresses one of the primary goals of \ac{socks}: enabling the seamless use of conventional Linux distributions on high-performance \acp{soc}, which provides distinct advantages over fully custom file systems. Leveraging proven binary packages from public repositories is faster and easier than compiling from source code. Using existing distributions also simplifies deployment and maintenance, as they integrate securely into existing network infrastructures and receive regular updates from public repositories, reducing the workload for developers and network administrators. However, distributions primarily developed for personal computers and servers also have limitations when used on embedded \acp{soc}, including limited adaptability and potentially unnecessary overhead. Standard distributions also do not offer binary packages for all \ac{cpu} instruction sets common in embedded devices. To account for these limitations, \ac{socks} also provides the flexibility to build a full Linux file system from source files, similar to Yocto and Buildroot. This is also realized using dedicated builders.

For clarity, the configuration of a \ac{socks} project shall be stored in a single file. It contains global settings that apply to all blocks, like the version of the Vivado toolset or the number of threads used for parallelized building. Additionally, it contains a section for every block that is contained within the \ac{soc} image. The block section contains a variety of block-specific settings and compulsory configurations, for instance, which builder is used to build the block. Condensing the project configuration in one place is a notable feature of \ac{socks} that distinguishes this framework from the established ones and simplifies the management of a project significantly.

The use of code forges like GitHub and GitLab in combination with \ac{cicd} practices is a widely used method to improve development efficiency and software quality. However, established \ac{soc} image development tools like Yocto are not ideally suited for \ac{cicd} workflows due to nested project structures, high storage requirements, and long build times. \ac{socks} was designed from the ground up for compatibility with \ac{cicd} and therefore offers a deeply embedded containerization feature that automatically builds each block in a separate container on a developer's local system. Each container has exactly the tools, toolchains, and dependencies that are required to build one specific \ac{soc} block or a group of blocks with similar requirements. The entire life cycle of the containers, from creating the images to starting and stopping them as needed, is covered by \ac{socks}. To maintain transparency, it must always be clear to the user when a container is being used. Developing in containers also brings benefits to local development by enabling flexible choice of the host system, improving reproducibility, and simplifying debugging. \ac{cicd} benefits from building blocks independently in different containers, because the execution of a number of smaller jobs is more flexible and often easier to handle compared to one big job. If desired, it is still easily possible to combine several individual containers into a single one for use in \ac{cicd}. To enable the transfer from local development to a \ac{cicd} pipeline, it must be possible to disable automated container use, and execute \ac{socks} itself in a container. This is required because it is common practice to run all jobs of a \ac{cicd} pipeline entirely in containers.

The most efficient development workflow can be achieved when local development is combined with \ac{cicd} pipelines. As soon as a pipeline exists, it is no longer necessary for a developer to build all blocks of an image locally. Only the blocks that are actually affected by the development work have to be built on the local machine, all others can be sourced pre-built from the pipeline. To enable this workflow, \ac{socks} is able to download the binary output files that are created when a block is built in a \ac{cicd} pipeline and integrate them into the local build process. To simplify this mechanism, \ac{socks} is capable of exporting and importing all output files for a block in a single compressed archive package.

\section{Implementation}

\acp{soc} are constantly evolving, and their areas of application can be highly specialized and diverse. As a result, any corresponding software and firmware build framework must be continuously adapted and improved to keep up with the hardware progress. Maintainability and extensibility are therefore decisive factors when selecting a suitable software architecture, designing the application, and choosing an appropriate programming language. For the latter, we have selected Python because it is accessible, widely used, and feature-rich. As an interpreted language, its source code does not need to be compiled, which accelerates development and lowers the barrier to participation in tool development and improvement. The comparatively high execution time is not a substantial drawback, given that computationally intensive tasks, such as compiling source code, are handled by specialized external programs. While the concept described in section \ref{sec:concept} is designed for a wide range of \ac{soc} architectures, it also dictates that each individual architecture must be explicitly implemented to ensure adequate support. Currently, the \ac{socks} implementation supports AMD's Zynq \ac{us+} and Versal families, and the Raspberry Pi 4 and 5.

\subsection{Software Architecture}
\label{sec:software_architecture}

\begin{figure}[b]
  \centering
  \includegraphics[width=\linewidth]{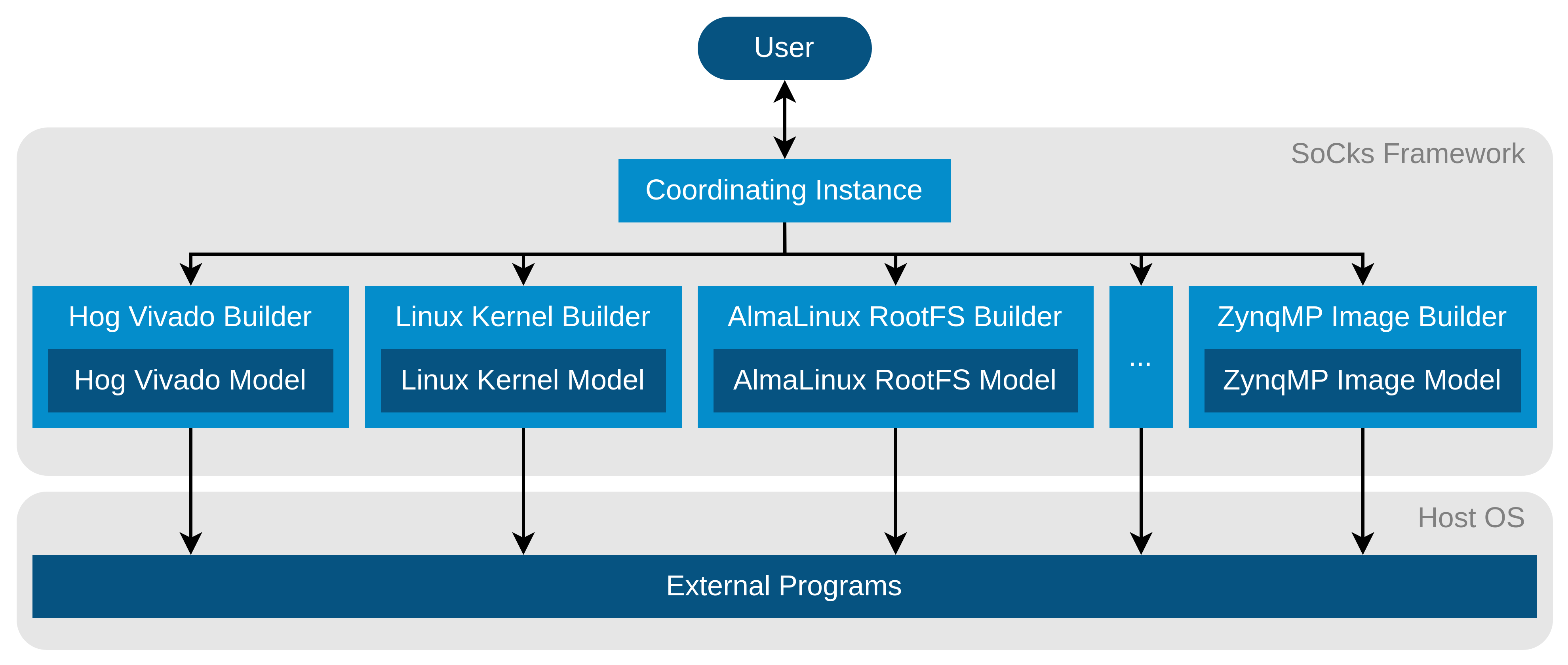}
  \caption{Architecture of the \ac{socks} Python application. The selection of builders shown is symbolic and not complete.}
  \label{fig:socks_architecture}
\end{figure}

Simplicity, modularity, maintainability, and extensibility were the main factors that guided the architectural design of the \ac{socks} framework. The foundation of the framework is the ``facade'' design pattern, a software design approach that provides a single object as the interface to a complex system \cite{Gamma1994}. In \ac{socks} this translates to a uniform user interface backed by a variable set of specialized subsystems, each dedicated to specific tasks. More precisely, an intuitive user interface is provided by a lightweight coordinating instance that manages program execution, while the main functionality is encapsulated in underlying subsystem modules, the so-called builders. Each builder is implemented as a Python class and specifically designed to create one component of the image, that is, one block. A dedicated Pydantic model is assigned to every builder, enabling it to parse, validate, and convert the corresponding section of the project configuration file into a Python-compatible data structure. Pydantic models are highly customizable, abstract representations of the expected data in the form of Python classes \cite{Pydantic}. A full introduction to \ac{socks} project configuration files follows in subsection \ref{sec:project_configuration}. The layout of the \ac{socks} Python application can be seen in Fig.~\ref{fig:socks_architecture}. In addition to the Python application itself, the \ac{socks} framework also contains supporting material such as container files and template packages with source files required by some builders. All of this is provided in one Python package but not explained in detail here for reasons of brevity.

\begin{figure}[t]
  \centering
  \includegraphics[width=\linewidth]{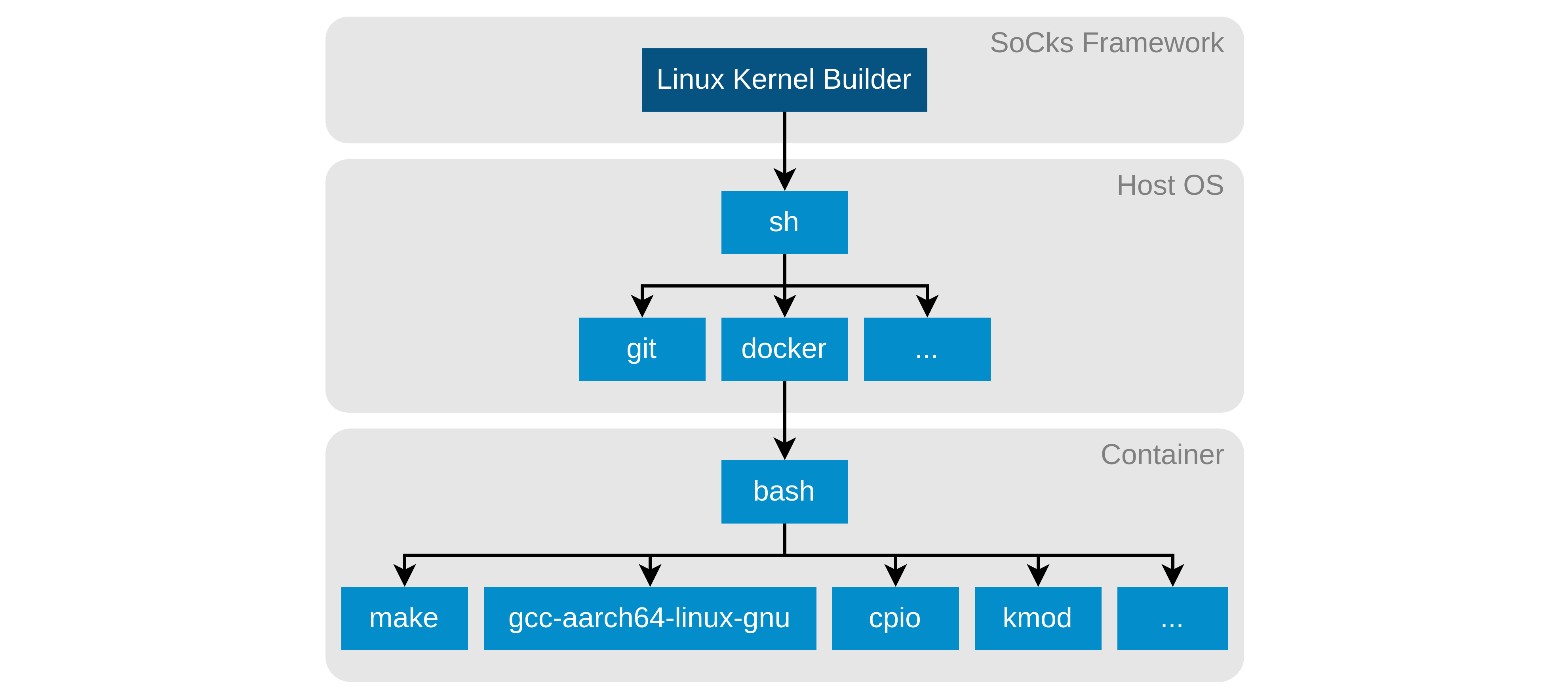}
  \caption{Access of a builder to programs on the host system and in the associated container. An external program is always accessed via a defined command interpreter (\ac{sh} or \ac{bash}), which ensures a uniform interface.}
  \label{fig:socks_external_tools}
\end{figure}

To keep the architecture of the framework simple, all builders are integrated via two standardized interfaces. One of them is dedicated to the coordinating instance, which uses it to control all actions of the builder. The second interface enables the builder to access external tools on the system via the command interpreter \ac{sh}. This basic POSIX shell-compatible interpreter was selected because it is available on all Unix systems. Fig.~\ref{fig:socks_external_tools} shows in detail how a builder accesses external tools. By default, \ac{socks} uses only a small set of tools on the host system. These are ``\mintinline{text}{git}'', GNU Core Utils like ``\mintinline{text}{hostname}'' and ``\mintinline{text}{id}'', as well as one of the containerization tools ``\mintinline{text}{docker}'' or ``\mintinline{text}{podman}''. All other tools required by the builder are usually provided in a container, specifically tailored for the individual blocks. The tools in the container are accessed via the command interpreter \ac{bash}. In contrast to \ac{sh}, \ac{bash} extends the POSIX shell syntax with several features that are beneficial when it comes to executing complex build-related commands in the container. If containerization support is disabled---for example, to use \ac{socks} itself in a \ac{cicd} pipeline container, as described in section \ref{sec:concept}---all required tools are expected to be available in the host environment. In relation to Fig.~\ref{fig:socks_external_tools}, this means effectively that the tools in sections ``Host OS'' and ``Container'' are merged on the host \ac{os} minus the containerization tool. Note that to ensure that all build-related commands are executed in the same way as in the container, \ac{socks} continues to use \ac{bash} in these cases.

\begin{figure}[t]
  \centering
  \includegraphics[width=\linewidth]{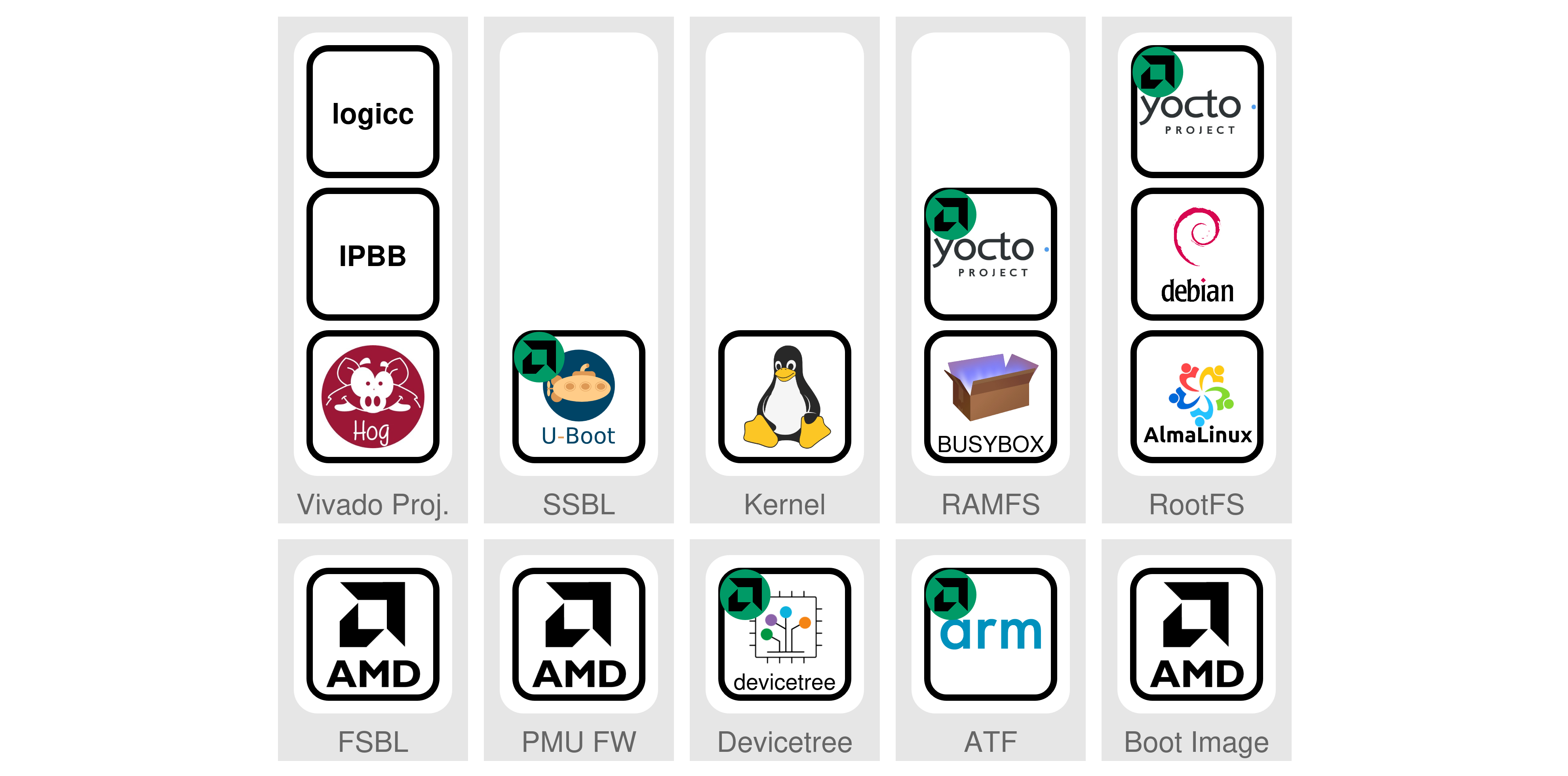}
  \caption{Overview of all builders currently provided by \ac{socks} for AMD Zynq \ac{us+} \ac{mpsoc} devices. Tiles with an AMD icon in the top left corner indicate that the builder is optimized for source code that has been adapted by the manufacturer for the target architecture.}
  \label{fig:socks_builders}
\end{figure}

As described in section \ref{sec:concept}, \ac{socks} uses different builders to represent different ways in which the content of a block can be implemented and built. However, it is not always necessary to create a new builder to support a new implementation of a block. If implementations differ just in the source code but utilize the same frameworks and toolchains, they can be supported by the same builder. For example, it is possible to build the Linux kernel provided by AMD and the official kernel from kernel.org with the same builder, whereas a Vivado project that utilizes the Hog framework and one that utilizes the \ac{ipbb} framework require different builders. Fig.~\ref{fig:socks_builders} provides an overview of all builders currently available in \ac{socks} for AMD Zynq \ac{us+} devices.

Although the various builder classes of the \ac{socks} framework focus on different aspects of the \ac{soc} image, they share much of their functionality. In addition to the basic interfaces just described, the functionality to import or export build artifacts of the blocks and the management of the container infrastructure are also shared. Following the principles of object-oriented software design, \ac{socks} provides an abstract builder base class that encapsulates all these functionalities together with the associated data. Beyond this, it is possible to identify several groups among all builders of \ac{socks} that share even more functionalities. Examples are builders that internally use AMD Xilinx development tools and builders that build file systems. The functionality shared in these groups is again encapsulated in abstract base classes. Similar to the builder classes, the various Pydantic model classes used to transfer the project configuration into a Python-compatible data structure also have functionality in common. For this reason, they also use common base classes to avoid code duplication and enforce a uniform layout of all model classes.


\subsection{Project Configuration}
\label{sec:project_configuration}

The project configuration file is the central element of every \ac{socks} project. We use the widely adopted YAML format to give users direct access to this file, avoiding the layer of abstraction that a configuration wizard would introduce. YAML is a popular data serialization format with implementations in many programming languages---including Python---that can be read and modified by humans with little training effort. Listing~\ref{lst:project_config_shortened} shows the basic structure of a \ac{socks} project configuration file.

\begin{listing}[t]
\begin{minted}
[
frame=single,
framesep=2mm,
fontsize=\footnotesize,
linenos,
numbersep=5pt,
xleftmargin=1em
]
{yaml}
import:
  - project-zynqmp-default.yml
project:
  type: "ZynqMP"
  name: "example-project"
external_tools:
  container_tool: "docker"
  xilinx:
    version: "2022.2"
    max_threads_vivado: 8
blocks:
  vivado:
    ...
  devicetree:
    ...
  rootfs:
    ...
  image:
    ...
\end{minted}
\caption{Abbreviated project configuration file for an AMD Zynq \ac{us+} \ac{mpsoc} image. Lines 1 to 10 contain the general project configuration, while the remaining lines contain block-specific settings. Since a default configuration for Zynq \ac{us+} devices is imported (line 2), this configuration file does not directly contain all settings required to build an image for this \ac{soc} architecture. This can be recognized, for example, by the fact that the ``\mintinline{text}{blocks}'' section (from line 11) does not contain subsections for all blocks required to build a complete Linux-based Zynq \ac{us+} image.}
  \label{lst:project_config_shortened}
\end{listing}

A \ac{socks} project configuration file consists of two sections, a general section and a section with block-specific information. The general section contains settings that cannot be assigned to a single block. This applies, for instance, to settings that are not processed by the builders of the blocks but by the coordinating instance of the framework itself. One example of this is the ``\mintinline{text}{import}'' section. Although the concept intends a single configuration file for every \ac{socks} project, there is the option to import external configuration files. This is necessary to outsource sensitive information, like usernames and encrypted passwords, into files that are ignored by version control. Additionally, it allows users to use predefined default project configuration files provided by the \ac{socks} framework for each supported \ac{soc} architecture. These default configuration files do not contain a complete project configuration, but they provide a starting point that can be further customized. The main purpose of the default configuration files is to simplify the creation of new \ac{socks} projects. However, they also reduce code duplication, as most projects for a given architecture may use the same settings in many of their parts. The usage of such a default project configuration file is depicted in line 2 of Listing~\ref{lst:project_config_shortened}. Furthermore, the general section also has settings that may impact multiple or even all of the blocks. Examples are the containerization tool or the version of the Vivado toolset to be used. The latter is particularly critical, as building several blocks with different versions of the Vivado toolset can lead to unpredictable incompatibilities.

Unlike the general section, the section with block-specific information is not a uniform unit but consists of a series of independent segments, each of which represents exactly one \ac{soc} block. After considering all files to be included, this block-specific section must be complete and contain every block that is used in the \ac{soc} image. To ensure strict separation of the individual blocks, the information stored in this section is not available to the builders of all blocks. Instead, each segment is only accessible to the builder of the block to which it refers and to the coordinating instance.

\begin{listing}[t]
\begin{minted}
[
frame=single,
framesep=2mm,
fontsize=\footnotesize,
linenos,
numbersep=5pt,
xleftmargin=1em
]
{yaml}
blocks:
  kernel:
    source: "build"
    builder: "ZynqMP_AMD_Kernel_Builder"
    project:
      build_srcs:
        source: "https://github.com/Xilinx/linux-xlnx.git"
        branch: "xilinx-v{{external_tools/xilinx/version}}"
      import_src: "https://serenity.web.cern.ch/.../kernel.tar.gz"
      add_build_info: false
      patches:
        - 0001-Add-build-information-to-proc.patch
      config_snippets:
        - disable-building-with-debug-info-to-reduce-size.cfg
    container:
      image: "kernel-builder-alma9"
      tag: "socks"
\end{minted}
\caption{Complete configuration section of the ``Linux Kernel'' block from a project configuration file for an AMD Zynq \ac{us+} \ac{mpsoc} image. This excerpt represents a possible extension of the project configuration in Listing~\ref{lst:project_config_shortened}.}
  \label{lst:kernel_config}
\end{listing}

Listing~\ref{lst:kernel_config} shows a complete configuration section of the ``Linux Kernel'' block. The layout of a block's configuration section depends on the builder used. However, some settings are mandatory for all blocks and builders. For example, the builder selection is a key part of this section and specified in ``\mintinline{text}{builder}''. Another example is the ``\mintinline{text}{source}'' parameter, which specifies whether the block is built locally or whether the build artifacts are imported from the location specified in ``\mintinline{text}{import_src}''. The remaining obligatory settings are in ``\mintinline{text}{container}'' and they determine which container is used to perform tasks related to this block. A characteristic of the block depicted in Listing~\ref{lst:kernel_config}---which is not shared by all blocks---is that its source files are located in a Git repository specified in ``\mintinline{text}{build_srcs}''. In this case, the repository is provided by AMD, which means that it is not possible to contribute project-specific changes to it. If only a few adjustments to the source code are needed, it is usually not reasonable to create a project-specific fork of the repository. Temporary changes can be made in the local repository, which \ac{socks} automatically clones into the project directory in preparation for building. If these changes are intended to be persistent, \ac{socks} can automatically create patches from local commits and integrate them into the project. To keep track of the order in which patches must be applied, \ac{socks} adds them to the list in ``\mintinline{text}{patches}''. Manual adjustments to this list are rarely needed. In case of a clean build, \ac{socks} clones the repository again and immediately applies all listed patches. One distinctive feature of the Linux kernel source code is that it includes a configuration based on the Kconfig language \cite{KconfigLanguage}. This configuration can be changed using tools such as ``\mintinline{text}{menuconfig}'' and is saved in the ``.config'' file in the root directory of the repository. Since the entire configuration is stored in a single file, patches are not a suitable method for recording changes. Instead, \ac{socks} uses so-called configuration snippet files. Similar to patches, they are listed in ``\mintinline{text}{config_snippets}'', and \ac{socks} automates the process of creating and applying them. A characteristic of \ac{socks} configuration files that enhances the YAML file format can be seen in line 8 of Listing~\ref{lst:kernel_config}. The double curly brackets represent a placeholder that allows one setting to be used to complement another. In this case, ``\mintinline{text}{external_tools/xilinx/version}'' is a reference to line 9 in Listing~\ref{lst:project_config_shortened}, which complements the string to ``\mintinline{text}{"xilinx-v2022.2"}''. This ensures that the branch of the kernel repository always matches the Vivado version used in the project.

\begin{listing}[t]
\begin{minted}
[
frame=single,
framesep=2mm,
fontsize=\footnotesize,
linenos,
numbersep=5pt,
xleftmargin=1em
]
{yaml}
blocks:
  image:
    source: "build"
    builder: "ZynqMP_AMD_Image_Builder"
    project:
      dependencies:
        atf: "temp/atf/output/bp_atf_*.tar.gz"
        devicetree: "temp/devicetree/output/bp_devicetree_*.tar.gz"
        fsbl: "temp/fsbl/output/bp_fsbl_*.tar.gz"
        kernel: "temp/kernel/output/bp_kernel_*.tar.gz"
        pmu_fw: "temp/pmu_fw/output/bp_pmu_fw_*.tar.gz"
        uboot: "temp/uboot/output/bp_uboot_*.tar.gz"
        vivado: "temp/vivado/output/bp_vivado_*.tar.gz"
        rootfs: "temp/rootfs/output/bp_rootfs_*.tar.gz"
    container:
      image: "amd-image-builder-alma9"
      tag: "socks"
\end{minted}
\caption{Complete configuration section of the ``Boot Image'' block from a project configuration file for an AMD Zynq \ac{us+} \ac{mpsoc} image. This excerpt contains information that was omitted in Listing~\ref{lst:project_config_shortened}.}
  \label{lst:image_config}
\end{listing}

Listing~\ref{lst:image_config} shows a valid configuration section of the ``Boot Image'' block. The main difference between the ``Boot Image'' block and the ``Linux Kernel'' block introduced before is that this block depends on the output of other blocks. In the configuration file, this requirement is expressed in the ``\mintinline{text}{dependencies}'' section, which contains paths to the output files of other blocks, so-called block packages. All paths in this section are relative to the \ac{socks} project folder, which means that they are independent of the exact location of the project. Typically, these relative paths are defined in a default project configuration file like ``project-zynqmp-default.yml'' and do not need to be adjusted by the user.

In \ac{socks}, the project configuration file is processed in three steps. First, all includes and placeholders are resolved to create a single data structure that contains the entire project configuration. Second, a project type-specific---i.e., \ac{soc} architecture-specific--- Pydantic model validates the general section of the configuration file and checks the presence of all required blocks. In the third and final step, all builders use their respective Pydantic models to validate the configuration section of their block. If an error occurs during one of the validation processes, the user is shown an error message specifying the exact location of the error in the configuration data. If the validation is successful, the user can display the complete and fully processed project configuration. Since \ac{socks} does not rely on a single configuration file as originally intended in section \ref{sec:concept}, this feature is essential for maintaining transparency and mitigating the complexity introduced by the include feature and the placeholders in the configuration files.

\subsection{User Interface}


Like the established frameworks Yocto and Buildroot, \ac{socks} uses a \ac{cli}. This simplifies the development and maintenance of the framework and enables its use in \ac{cicd} pipelines. With \ac{socks}, it is possible to build a complete \ac{soc} image or single blocks with just one shell instruction. Listing~\ref{lst:socks_build_cmd} shows an example. Interacting with one or more blocks always requires \ac{socks} to be executed with the following two parameters.

\begin{listing}[b]
\begin{minted}
[
frame=single,
framesep=2mm,
fontsize=\footnotesize,
linenos,
numbersep=5pt,
xleftmargin=1em
]
{bash}
$ socks kernel build
\end{minted}
\caption{\ac{bash} instruction to build the ``Linux Kernel'' block of a \ac{socks} project.}
  \label{lst:socks_build_cmd}
\end{listing}

\begin{description}
   \item[Block] specifies the block to be operated on by its block ID. In the example shown in Listing~\ref{lst:socks_build_cmd}, this is the block for building the Linux \textit{kernel}. However, it is also possible to use the keyword \textit{all} instead of a block ID to target all blocks of the image at once.
   \item[Command] contains the command to be applied to the specified block or blocks. The command \textit{build} used in Listing~\ref{lst:socks_build_cmd} generates the output products of the specified block. The set of available commands depends on the block---more precisely, on its builder---but they can always be grouped into four categories: building, configuring, debugging, and cleaning.
\end{description}

\begin{listing}[b]
\begin{minted}
[
frame=single,
framesep=2mm,
fontsize=\footnotesize,
linenos,
numbersep=5pt,
xleftmargin=1em
]
{text}
$ socks kernel --help
usage: socks kernel [-h] [-g]
                    {prepare, build, clean, create-patches, create-cfg-snippet,
                    start-container, menucfg}
                    ...

Build the official AMD/Xilinx version of the Linux Kernel for ZynqMP devices

options:
  -h, --help            show this help message and exit
  -g, --group           Interact not only with the specified block, but also with all blocks
                        on which this block depends.

commands:
  {prepare,build,clean,create-patches,create-cfg-snippet,start-container,menucfg}
    prepare             Performs all the preparatory steps to prepare this block for
                        building, but does not build it.
    build               Builds this block.
    clean               Deletes all generated files of this block.
    create-patches      Uses the commited changes in this block's repo to create patch files.
    create-cfg-snippet  Creates a configuration snippet from the changes in the .config file
                        in this block's repo.
    start-container     Starts the container image of this block in an interactive session.
    menucfg             Opens the menuconfig tool to enable interactive configuration of the
                        project in this block.
\end{minted}
\caption{Help text of \ac{socks} for the ``Linux Kernel'' block.}
  \label{lst:socks_help}
\end{listing}

In addition to the two parameters, it is possible to specify options, which are indicated by one or two preceding dashes (e.g., ``\mintinline{text}{-h}'' or ``\mintinline{text}{--help}''). The ``\mintinline{text}{--help}'' option can be specified directly after the executable ``\mintinline{text}{socks}'' or after any of the parameters to display the respective help text. Listing~\ref{lst:socks_help} shows the help text of the first parameter, in this case specifying the block to build the Linux kernel. The content of these help texts is largely determined by the project configuration. In Listing.~\ref{lst:socks_help}, this can be seen in the commands for interacting with the block. These commands depend on the block's builder selected in the project configuration. To address these dependencies, the help texts of \ac{socks} are generated at runtime based on the specific user project.

\subsection{Image Creation Process}

\begin{figure}[b]
  \centering
  \includegraphics[width=\linewidth]{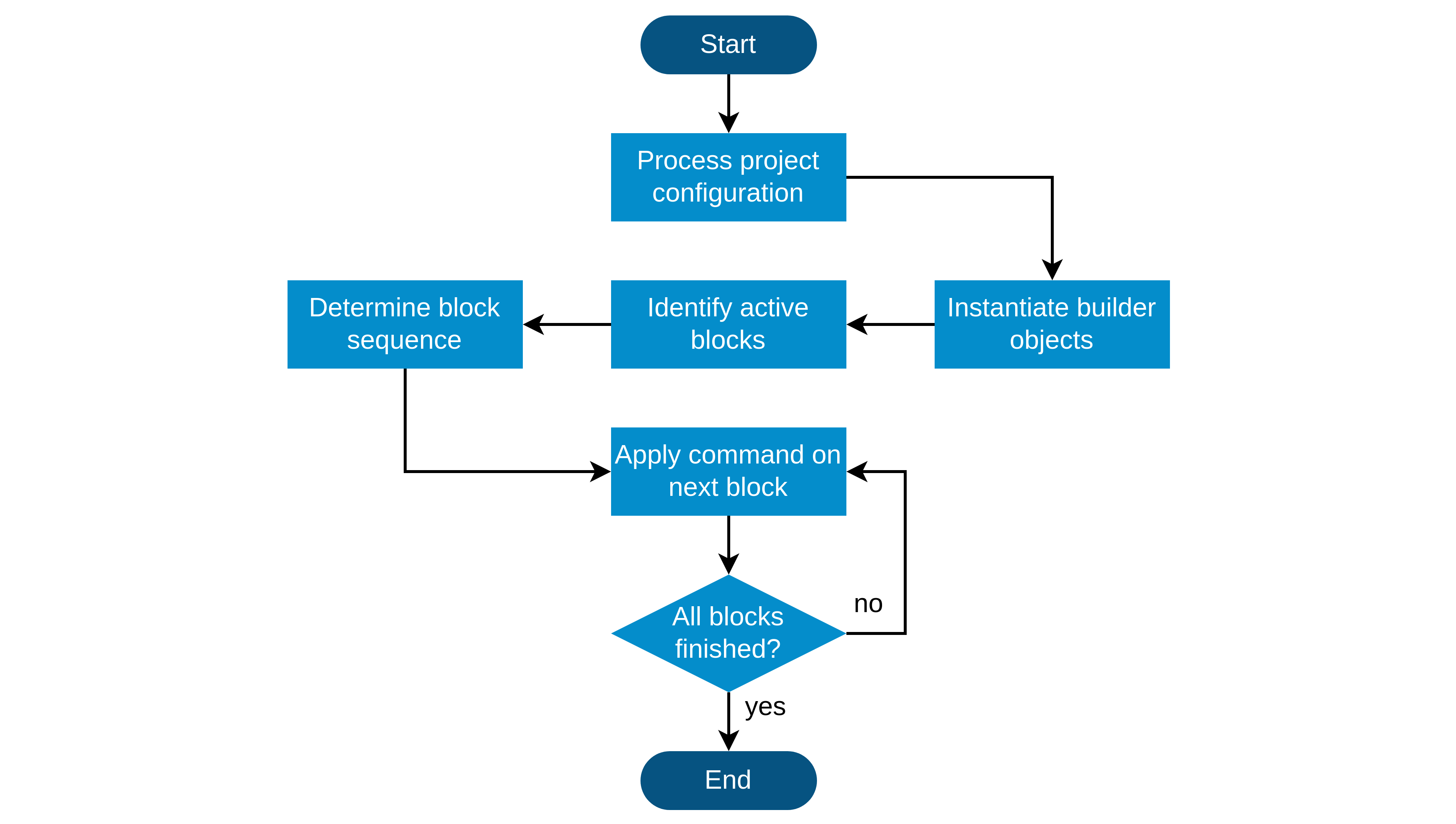}
  \caption{Process used by the coordinating instance to apply a command to one or more blocks.}
  \label{fig:build_flowchart}
\end{figure}

To build a bootable \ac{soc} image from the user project, \ac{socks} uses a multi-step process, which is shown in Fig.~\ref{fig:build_flowchart}. This high-level sequence forms the core of the coordinating instance and is used not only to build complete images but also to apply any command to any number of blocks. In the following section, the command ``\mintinline{text}{build}'' will be used as an example. The process always begins with the basis of every \ac{socks} project, the project configuration file, being read and processed. This information is then used to dynamically instantiate all specified builders. Dynamic instantiation means that \ac{socks} does not require a fixed set of builder classes but uses only those specified in the project configuration file. This simplifies the modular extension of the framework. 

The next step is to identify all active builders. These are builders that are addressed directly or via transitive dependencies in the call to \ac{socks}. If the call contains the keyword ``all'', for instance to build a complete image, all builders are addressed. However, if only the root file system and all blocks on which it depends are to be built, Fig.~\ref{fig:partitioned_image} shows that only the ``Root File System'' block itself, the ``Linux Kernel'' block, the ``Devicetree'' block, and the ``Vivado Project'' block are addressed. Although this example shows that not all builders of a project may be actively used for a specific action, it is still required to initialize all of them, because the builders are used to ensure a complete and valid project configuration. 

Once the subset of active builders is known, they are arranged in the sequence in which they will be used. This sequence results from the dependencies between the blocks and from the command that is applied to them. When building, all dependencies of a block are built prior to the block itself. Cleaning, for instance, should be carried out in reverse order so that the most fundamental blocks are not cleaned first. This difference can be important if the user decides to cancel the cleaning process. In this case, \ac{socks} immediately stops cleaning and preserves the files of the remaining blocks that have not yet been processed.

Once the sequence has been defined, the command specified by the user is applied to the builders one after the other. This sequential approach is suitable for a lightweight tool, as it is easy to handle and can be tracked by the user at runtime. A significant performance disadvantage is also not expected, as most builders use internal parallelization, for example, via tools such as Make, Ninja, or Vivado. By default, \ac{socks} instructs these tools to use all cores available on the host system. However, it is also possible to set a lower number in the project configuration file. Once all blocks have been processed, the \ac{socks} application is closed. If \ac{socks} was called to build a complete image, the results can now be copied to the boot medium of the target \ac{soc} and executed from there.

To be applicable to any \ac{soc} architecture, the high-level process described so far is strongly decoupled from the actual construction of an \ac{soc} image. The architecture-specific processes are implemented in the builders and can be freely designed. However, to be interchangeable, all builders have standardized interfaces for data flow. These are not to be confused with the software interfaces via which builders are integrated into the \ac{socks} framework as described in subsection \ref{sec:software_architecture}. 

\begin{figure}[b]
  \centering
  \includegraphics[width=\linewidth]{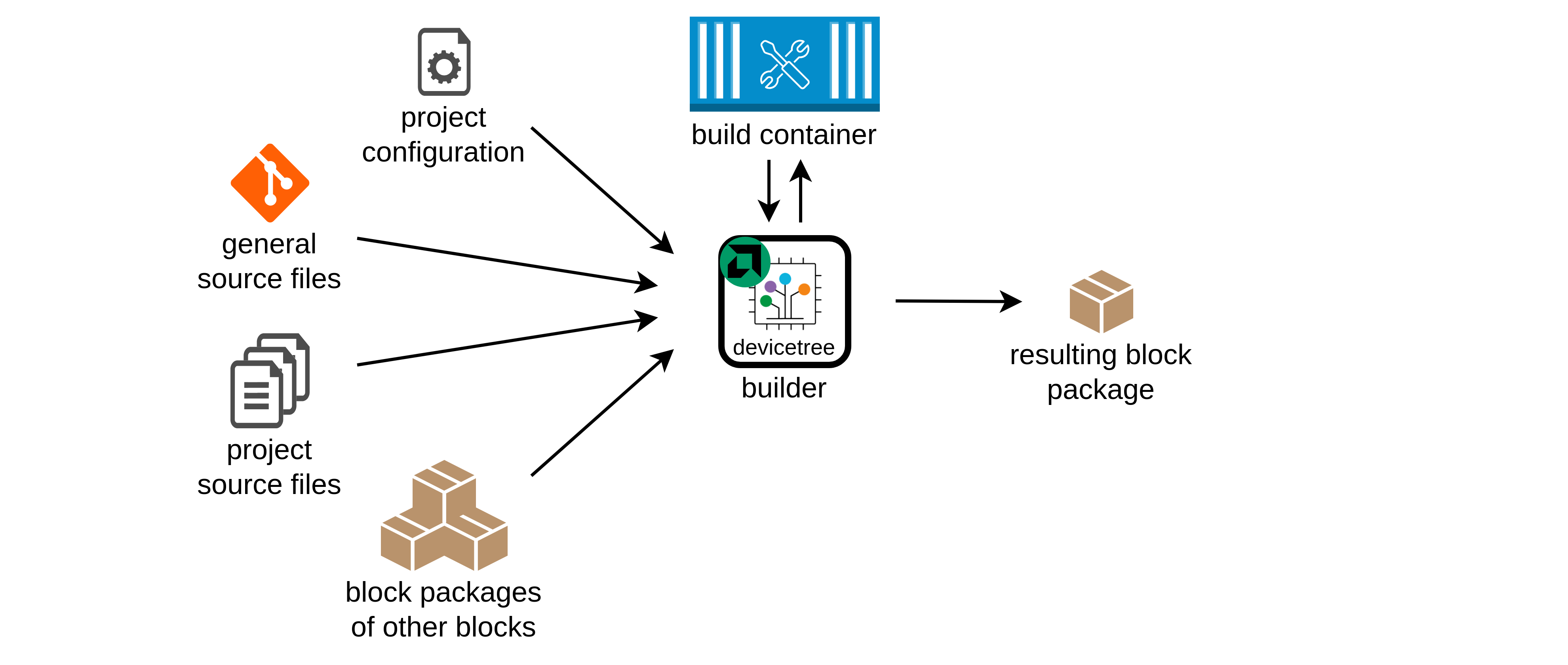}
  \caption{Data flow interfaces of a \ac{socks} builder. The devicetree builder shown in this example uses all possible data inputs, but some builders only use a subset of them.}
  \label{fig:block_build}
\end{figure}

Fig.~\ref{fig:block_build} provides an overview of the data flow interfaces that a builder can use. The four possible data sources can be seen on the left. Apart from the ``project configuration'', these interfaces are optional. A usually self-contained software or \ac{fpga} firmware project can be specified as ``general source files''. This project must be provided as a Git repository, either via a \ac{url} or as a local path. One example is the source repository of the Linux kernel. In addition, ``project source files'' can be specified, which, in contrast to the ``general source files'', directly relate to the specific user project and must therefore be available locally. Examples are patches for the ``general source files'' or template files that define the layout of a binary boot file. Finally, there are so-called ``block packages of other blocks''. These are compressed tar archives specified in the project configuration under the ``\mintinline{text}{dependencies}'' section of the block, as can be seen in Listing~\ref{lst:image_config}. Block packages contain build artifacts from other blocks that are further processed by the builder of this block. Although tar archives are a flexible format for transferring data from one block to another, the required content can be specified to ensure reliable information transfer. For this purpose, the builders have a mechanism to validate the content of the received block packages depending on the emitting block. For example, the Zynq \ac{us+} devicetree builder enforces that the block package it receives from the ``Vivado Project'' block must contain an \ac{xsa} file. In contrast, file system builders can integrate kernel modules into the file system if they find them in a ``Linux Kernel'' block package, but they do not enforce their presence, as the Linux kernel can also be configured not to use external modules.

Fig.~\ref{fig:block_build} also depicts the option offered by \ac{socks} to use containers to provide blocks with suitable build environments. Depending on the project configuration, blocks can either use their own container image or share one with other blocks. If required, the image is automatically created at build time by the builder using the container files included in the \ac{socks} framework. The builder then uses the container image autonomously, for example, to generate the output products of the block. Once all building processes have been successfully completed, the builder provides its data output by packaging the products into a block package so they can be further processed by another builder if required. For debugging purposes, the user can also enter the containerized build environment manually.

Incremental building can significantly accelerate build processes and is therefore widely adopted in modern build tools such as GNU Make, Ninja, Rust's Cargo, and Yocto \cite{GNUMakeManual, NinjaManual, Rust1_24, YoctoSharedStateCache}. This strategy is based on the fact that not all output products must be recreated during each build, but only those whose sources have changed. For an efficient implementation, it is essential that checking whether a component needs to be rebuilt takes significantly less time than building the component itself. \ac{socks} uses three different approaches to enable incremental builds.

\begin{description}
   \item[Timestamps] The most fundamental method used in \ac{socks} to check whether output files need to be recreated is to compare timestamps. To do this, the timestamp of the last modified source file is compared with the timestamp of the last modified output file. If the source file is newer, the component must be rebuilt. Simplicity is the major advantage of this method. However, for directories with many files, it can take a long time to find the last modified file, which is a significant disadvantage in these cases.
   \item[Event log] When the success of a stage cannot be verified based on specific output files---for example, because it is not possible to predict which files will be generated or because they are not easily accessible, as is the case when building a Docker container---the aforementioned timestamp-based method is not feasible. In such cases, \ac{socks} uses event log files. These files are in \ac{csv} format and contain timestamps of successful build stages, each with a unique ID that identifies the respective stage. These timestamps are compared with the timestamp of the last-modified source files to determine if a rebuild is required.
   \item[Checksums] When importing files from an archive, timestamps are not a reliable source of information, as the timestamps of the extracted files are retained from before they were packed. Therefore, \ac{socks} uses checksums to verify whether a provided archive has already been imported.
   \item[Configuration comparison] One limitation of the three methods mentioned above is that the smallest granularity they can capture is at the level of files. For the project configuration file, this is a problem because a minor change would mean that the entire user project must be rebuilt. To prevent this, each builder saves a copy of the project configuration it used once the build process was successful. In a subsequent build process, this copy is used to detect individual changes in the project configuration and decide whether a rebuild is required.
\end{description}

Often, the incremental build mechanisms implemented in \ac{socks} are only the first layer. Some of the build tools used by \ac{socks}---such as GNU Make and Yocto---use incremental build mechanisms themselves. These tools have additional information about the components to be built and can therefore decide more precisely which parts actually need to be rebuilt. Nevertheless, it is beneficial to use the incremental build features of \ac{socks} in these cases as well, as these features can save overhead, such as avoiding unnecessary container startups.

\section{Distributed development}

\begin{figure}[b]
  \centering
  \includegraphics[width=0.90\linewidth]{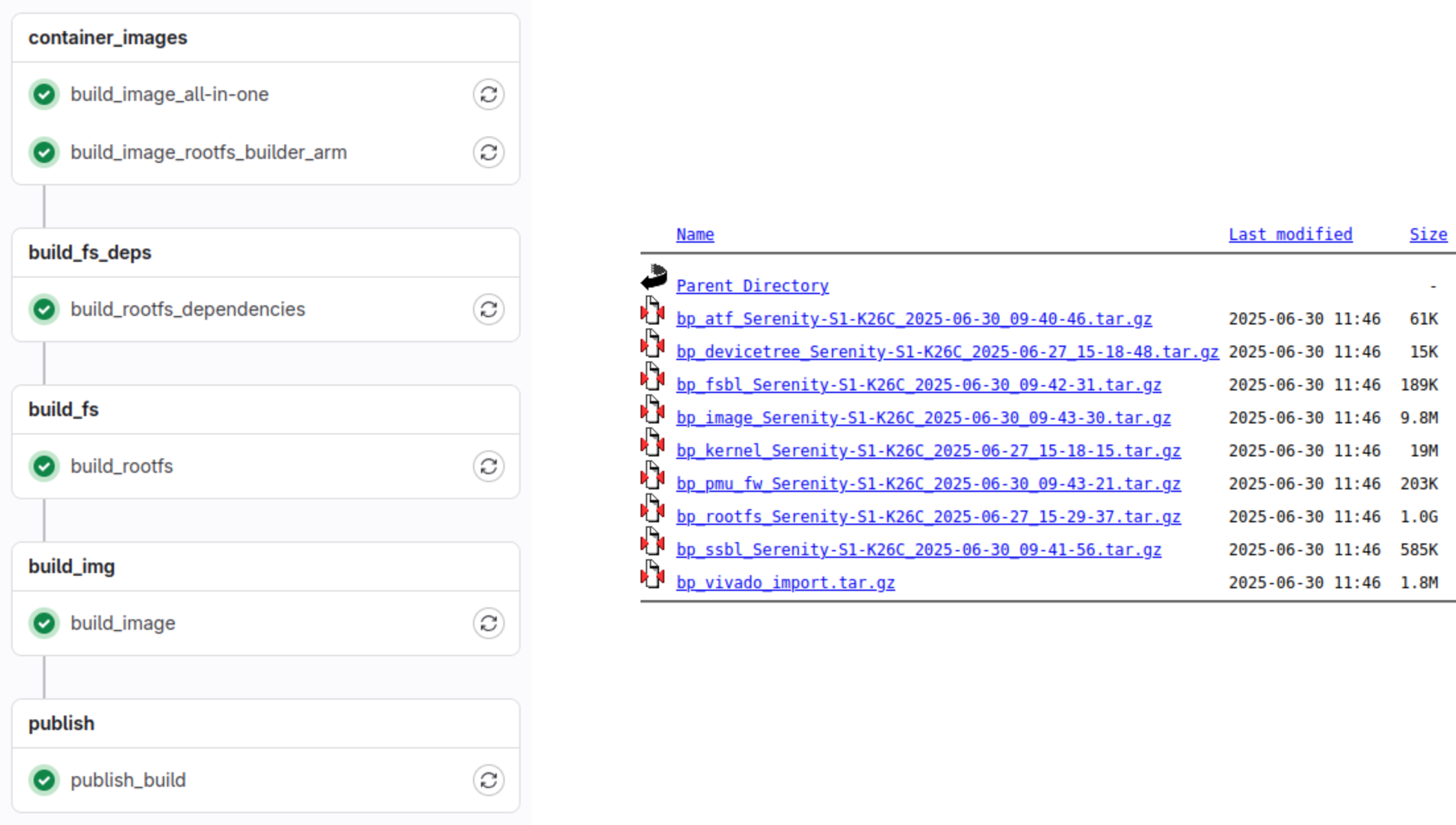}
  \caption{The left-hand side shows a GitLab pipeline that builds a complete image for an AMD Zynq \ac{us+} \ac{mpsoc}. The first stage, ``container images'', builds the container images that are used for the subsequent stages. In this example, not every block is built in an independent job. Instead, the blocks are built in three groups. The first group ``build\_fs\_deps'' contains all blocks on which the root file system depends, the second group ``build\_fs'' contains only the ``Root File System'' block itself, and the third and last group ``build\_img'' contains all remaining blocks. Finally, the concluding stage ``publish'' uploads all build artifacts to a server. The concept behind this pipeline is to isolate the ``Root File System'' block so that it can be built on an AArch64 system, while all other blocks are built on an x86-64 system. This eliminates the need to emulate the AArch64 architecture when building the root file system. The files that the pipeline has uploaded to the server are shown on the right-hand side. The ``Vivado Project'' block package name differs from the standard naming scheme because it is imported from a dedicated Vivado pipeline rather than being built in this pipeline.}
  \label{fig:gitlab_pipeline}
\end{figure}

Modern high-performance \ac{soc} devices are complex systems whose images are usually created by a team of developers. In the scientific community, it is common that not all members of this team are in the same location and therefore rely on distributed development practices. \ac{socks} was developed in such an environment and was therefore designed to support distributed development from the very beginning. The central element of this workflow is the use of \ac{cicd} pipelines that build the complete \ac{soc} image and publish all block packages on a server that is accessible to all team members. An example of such a pipeline is shown in Fig.~\ref{fig:gitlab_pipeline}.

Team members working on the \ac{fpga} firmware implementation in Vivado---one of the early blocks in the \ac{soc} image build chain that does not have any dependencies on other blocks---can either build the full \ac{soc} image on their local machine, or they can build and test their implementation in Vivado independently and then push the modified source files to the Git repository, where the \ac{cicd} pipeline builds the full \ac{soc} image. The finished \ac{soc} image can then be downloaded and tested by any team member.

For team members who are working on a block further down in the build sequence, there are additional advantages. Root file system developers, for instance, can download all block packages except the boot image and the root file system itself from the server that holds the build artifacts from the \ac{cicd} pipeline (see dependencies in Fig.~\ref{fig:partitioned_image}). Then, they can adapt the \ac{socks} project configuration file so that these blocks are not built locally but instead imported from the provided block packages. If the server supports downloading individual files via a \ac{url}, they can also modify the project configuration so that \ac{socks} automatically downloads the files. This ensures that the block packages are automatically updated when new versions are available. If a root file system developer wants to build a complete \ac{soc} image locally for testing, \ac{socks} will effectively only build the ``Root File System'' and the ``Boot Image'' block on the local machine. This can save a considerable amount of time and resources. It can also eliminate the need for all developers to have all the development tools required for the full \ac{soc} image available, including any necessary licenses.

The most significant advantage arises for team members developing software to be incorporated in the root file system if the \ac{socks} project uses a conventional distribution, such as Debian or AlmaLinux. In this case, these team members can almost fully decouple their development workflow from the \ac{soc} context and develop Debian or AlmaLinux packages in much the same way as they would for a desktop PC. The main remaining difference is cross-compilation, which may be required depending on the development environment. For testing, these packages can then be installed to the \ac{soc} image at runtime via the respective package manager. \ac{socks} provides three methods to install such user-defined packages at build time: a package that is locally available on the development machine can be installed via a path specified in the project configuration file; a package hosted on a server can be installed via a \ac{url}; and packages can also be installed directly from self-hosted repositories of the respective package manager.


\section{Performance and Comparison}


Depending on the architecture of the \ac{soc} and the associated \ac{soc} image, resource consumption and duration of the build process can vary greatly. All measurement results presented in this section serve comparative purposes and are not representative of using the framework in general. To obtain comparable results, an equivalent image was implemented using the \ac{socks} and the Yocto framework. Yocto was chosen as the reference framework because AMD recommends it for their high-performance \acp{soc} \cite{PetaLinux2Prod}. Given the vast number of configuration options Yocto provides, universally valid performance measurements are difficult to achieve. To facilitate reproducibility, the Yocto Project was configured according to AMD’s recommendations using their official layers. Only necessary, project-specific adjustments were made. Although Yocto offers caching mechanisms such as the shared state cache and user-defined package repositories, these mechanisms were not used because it is not in the default configuration \cite{YoctoSharedStateCache, YoctoPackageDocu}. Furthermore, the effectiveness of these mechanisms is not consistent, and they must be hosted locally, which creates additional effort for developers. The test image is based on version 2022.2 of the AMD toolset and targets the Zynq \ac{us+} \ac{mpsoc} on the DTS100G card \cite{Muscheid_2023}. The implementations in both frameworks use logicc to build the Vivado project. As it is officially not intended to build the Vivado project with Yocto, a custom recipe was added to enable this. According to the recommended choice for local development, the \ac{socks} project was configured to use Docker containers. Furthermore, a Debian root file system was used in the \ac{socks} project, unless mentioned otherwise. All measurements were carried out on the same AlmaLinux 8.10 test system with an Intel Core i9 14900K, 128\,GB of DDR5 memory, and a 2\,TB Samsung 990 EVO \ac{nvme} \ac{ssd}. AlmaLinux 8 was chosen as the \ac{os}, because it is binary-compatible with \ac{rhel} 8, which is officially supported by the AMD toolset version 2022.2 and the corresponding Yocto version ``Honister'' \cite{VivadoReleaseNotes, PetalinuxRefGuide, YoctoSystemRequirements}. We chose the 2022.2 version of the AMD toolset for these tests because this is the version currently used in most of our long-term projects. However, the implementation of \ac{socks} is designed to be as independent as possible from the version of the AMD toolset and has also been successfully tested with versions 2020.2 and 2024.2.

\begin{figure}[t]
  \centering
  \includegraphics[width=\linewidth]{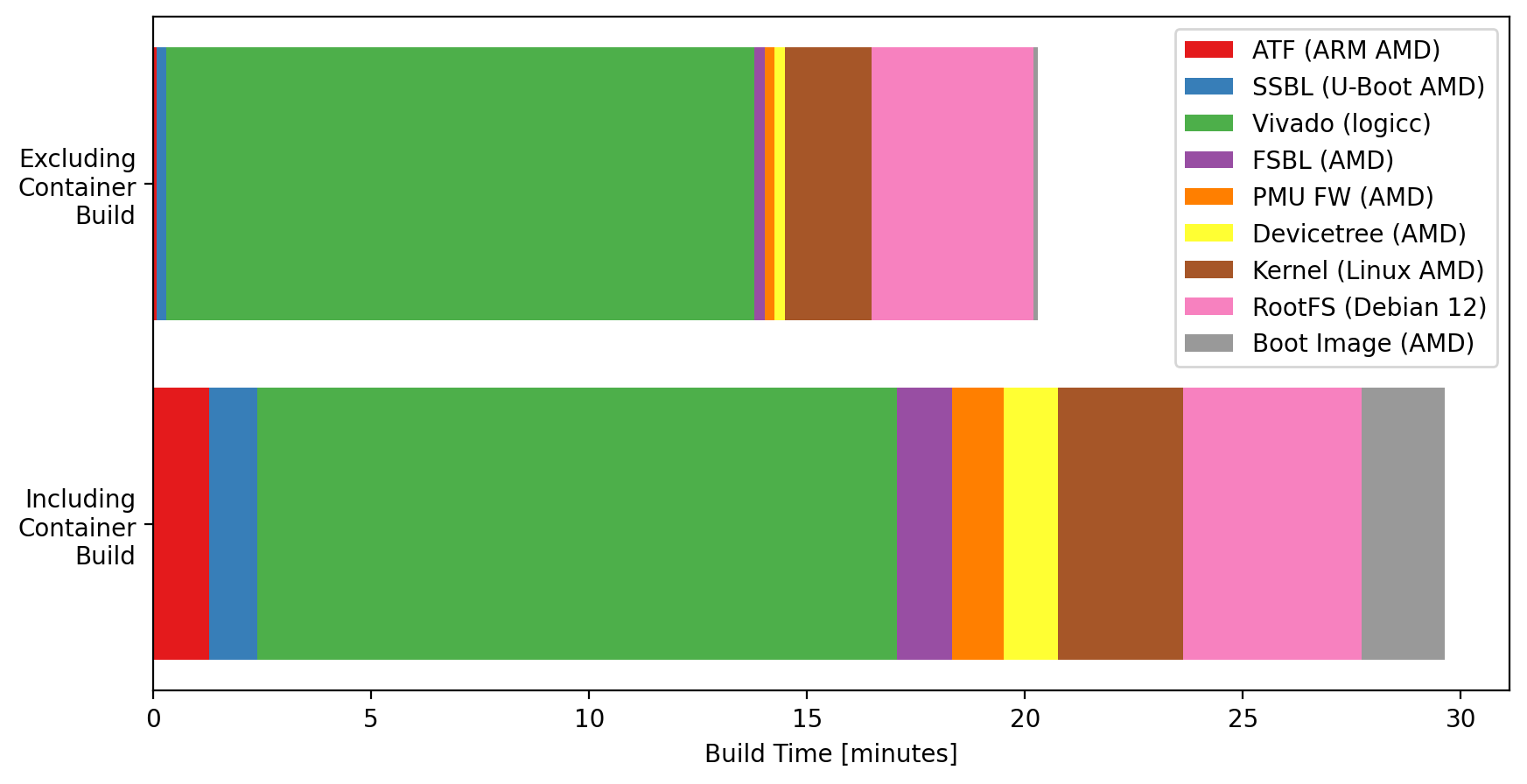}
  \caption{Overview of the build time of the individual components of a \ac{socks} image. All blocks were built individually, so this diagram does not give a representative statement on the build time of a complete image. All measurements were carried out five times and averaged.}
  \label{fig:comp_build_comp}
\end{figure}

\begin{table}[t]
    \centering
    \caption{Resource utilization of the Vivado project used for all tests. The target device is a xczu19eg-ffvc1760-2-e.}
    \label{tab:vivado_prj}
    \begin{tabular}{cccc}
        \toprule
        Resource&Utilization&Available&Utilization \%\\
        \midrule
        LUT&20995&522720&4.02\\
        FF&29557&1045440&2.83\\
        BRAM&489.5&984&49.75\\
        DSP&3&1968&0.15\\
        \bottomrule
    \end{tabular}
\end{table}

\begin{figure}[b]
  \centering
  \includegraphics[width=\linewidth]{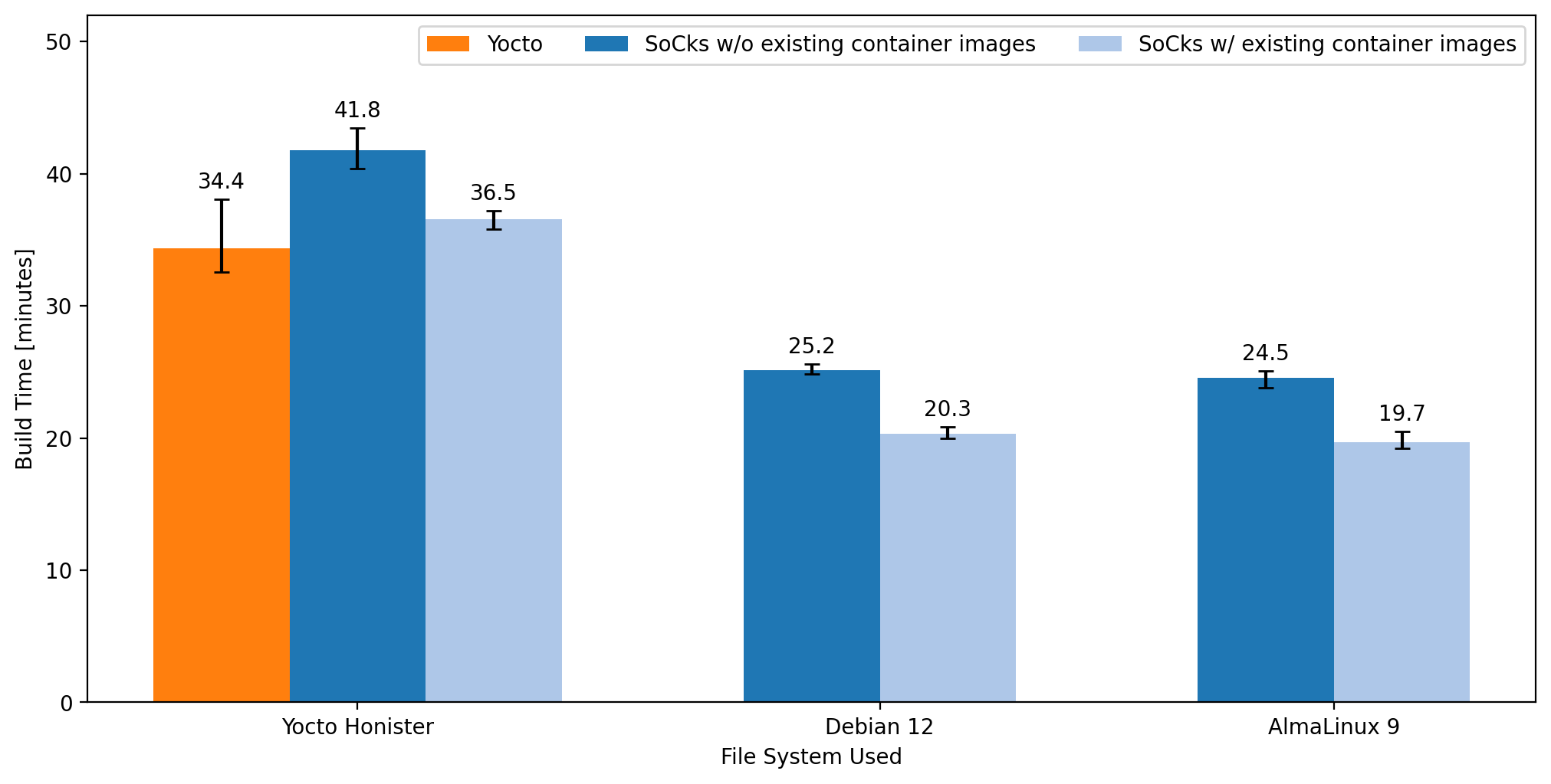}
  \caption{Comparison of the build time of different complete \ac{soc} image projects. The \ac{socks} images differ only in the root file system used. The Yocto framework is not designed to build images with file systems of conventional distribution. Therefore, the Yocto project was built exclusively with the Yocto Honister file system. All measurements were carried out five times and averaged. The error bars indicate the range of the measurement results.}
  \label{fig:image_build_comp}
\end{figure}

The various \ac{soc} blocks that form a \ac{socks} image do not contribute equally to the overall build time. Most time is generally taken by the ``Vivado Project'' block, as the example in Fig.~\ref{fig:comp_build_comp} shows. Since this block's build time depends heavily on the specific Vivado project, Table~\ref{tab:vivado_prj} gives an impression of the project size based on resource utilization. A major cause of the long build time of the Vivado project is that large parts of the creation process of the \ac{fpga} bitfile are not executed in parallel and therefore do not use the \ac{cpu} efficiently. The other \ac{soc} blocks that contribute above average to the construction time are the ``Linux Kernel'' and the ``Root File System''. These are the two largest software components in the \ac{soc} image, which is reflected twofold in their build time. The source files for the Linux kernel are several gigabytes in size, which is why not only compiling but also downloading the sources is a significant portion of the build time of the block. Similarly, the Debian file system used in the example in Fig.~\ref{fig:comp_build_comp} is constructed from a large number of individual packages that also need to be downloaded. In comparison with the aforementioned three blocks, the build time of the remaining \ac{soc} blocks is almost negligible. However, this changes if \ac{socks} has to build the containers before it can build the blocks themselves. The overhead introduced by these processes can be significant, especially for blocks with a short build time. The bar ``Including Container Build'' in Fig.~\ref{fig:comp_build_comp} shows the worst-case scenario in this respect, because all blocks were built independently of each other, each on a clean system without any existing container images. If an \ac{soc} image is built on a clean system as a whole, the build time of container images later in the chain benefits from already existing containers, because they have common layers that are automatically reused by Docker. Furthermore, there are blocks like ``FSBL'' and ``PMU FW'' that typically use the same container image. Since the container images are managed by the containerization tool, they are also independent of the specific \ac{socks} project. This means that a container image only needs to be built if it is used for the first time on a system or if \ac{socks} was updated. A comparison of Fig.~\ref{fig:comp_build_comp} with the corresponding bars in Fig.~\ref{fig:image_build_comp} shows the difference between building all blocks independently and creating the \ac{soc} image as a whole.

%

Fig.~\ref{fig:image_build_comp} also shows that when using Debian or AlmaLinux, \ac{socks} can build a complete \ac{soc} image considerably faster than Yocto, even if all containers need to be built. The only time \ac{socks} is slower than Yocto is when the Yocto file system is used. This is because \ac{socks} uses Yocto in a container. In this case, \ac{socks} depends on Yocto's performance, and the redundant infrastructure of both build frameworks adds up, resulting in poor performance. Comparing projects that rely on different file systems may seem unequal, but the ability to use file systems from conventional distributions on images for embedded \acp{soc} is a unique feature of \ac{socks} that neither Yocto nor Buildroot support. Therefore, a direct comparison with file systems of regular distributions is not possible. If the \ac{socks} project uses a Debian or AlmaLinux root file system and the required containers already exist on the system, building the project is, on average, about 40\,\% faster compared to Yocto. The primary reason is that Yocto builds all components for the file system and even many host build tools from source. This enables great flexibility but is also very inefficient. In contrast, \ac{socks} downloads the precompiled packages for the Debian or AlmaLinux root file system, and the required toolchains for the host systems are, if possible, also downloaded in compiled form when the corresponding containers are created. The test system used for the measurements has a powerful \ac{cpu} and Yocto uses parallelization efficiently to utilize all available cores, but on less powerful systems the difference between Yocto and \ac{socks} is even larger, because compilation time increases, while the download speed of precompiled files is unaffected. Tests with the same \ac{soc} image projects on an AlmaLinux 8.10 system with older hardware---specifically, an Intel Core i7 6700, 32\,GB of DDR3 memory, and a 500\,GB Samsung 850 EVO \ac{sata} \ac{ssd}---showed that it takes about three times as long to build the Yocto project as the equivalent \ac{socks} project with a Debian file system ($\bar{x} \approx$ 150 minutes vs. $\bar{x} \approx$ 49 minutes, if the required containers already exist).

\begin{figure}[b]
  \centering
  \includegraphics[width=\linewidth]{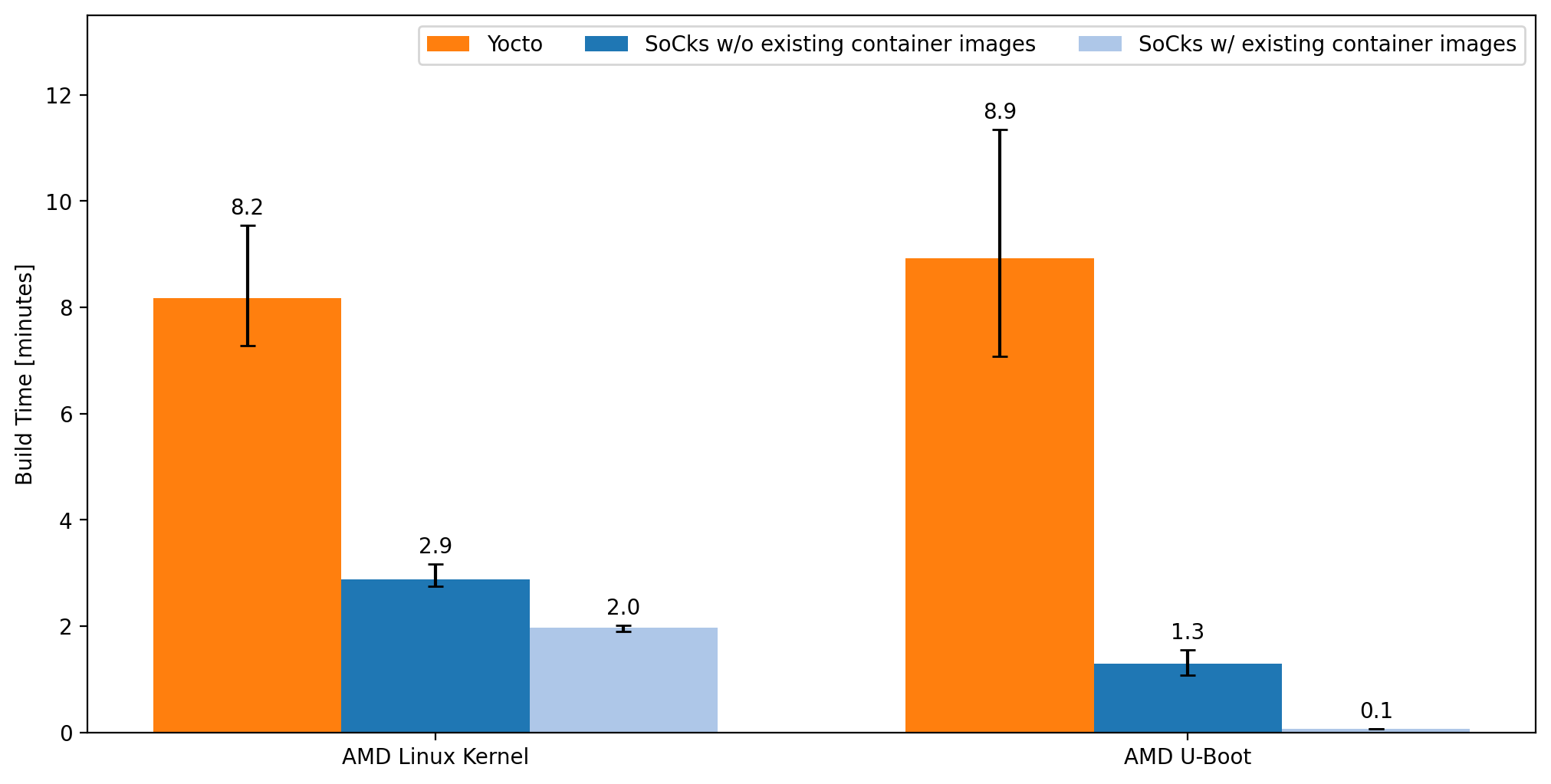}
  \caption{Comparison of the build time of individual components of an \ac{soc} image. All measurements were carried out five times and averaged. The error bars indicate the range of the measurement results.}
  \label{fig:comp_build}
\end{figure}

If \ac{socks} and Yocto are used to build a single component, the difference in build time can be even larger, as shown in Fig.~\ref{fig:comp_build}. The overhead of Yocto in this case is mainly caused by the fact that it builds the required toolchain locally. If a complete image is to be built, this overhead is spread across several components that use the same toolchain, but if just one component is to be built, the practice of building the toolchain locally is highly inefficient. The complex web of dependencies in a Yocto project can also add overhead. Although no custom recipes were used in the tests carried out for Fig.~\ref{fig:comp_build}, it cannot be ruled out that some of the recipes used may be inefficiently designed and include superfluous dependencies. The error bars in Fig.~\ref{fig:comp_build} indicate another weakness of Yocto. Across all measurements, it was observed that Yocto's download speeds are very inconsistent. This was not observed with \ac{socks}, so the problem is most likely not due to the network connection of the test system but to the servers that Yocto uses to download the data.

\begin{figure}[t]
  \centering
  \includegraphics[width=\linewidth]{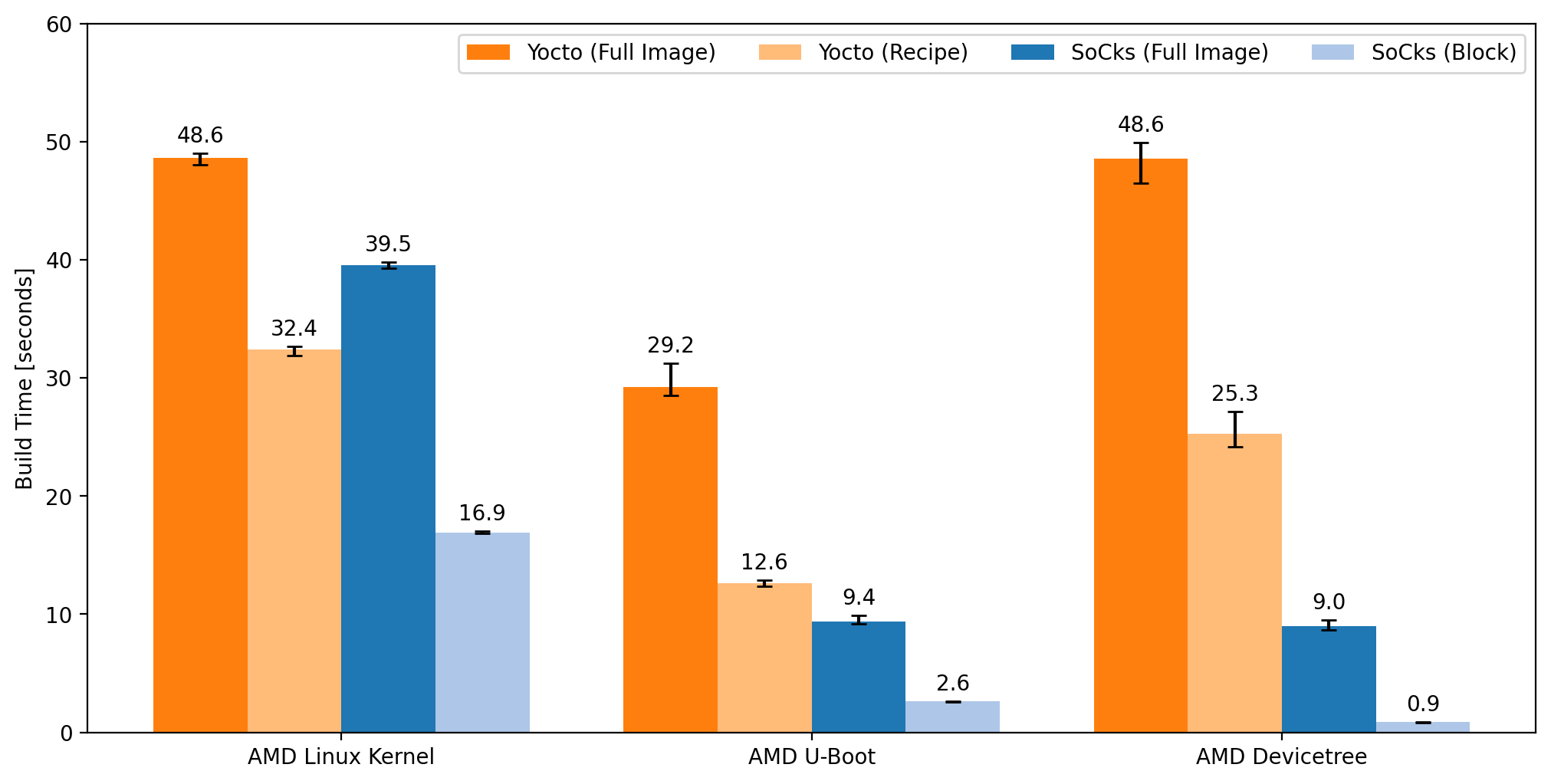}
  \caption{Comparison of the rebuild time after a source file for the respective component has been edited. In both the Linux kernel and U-Boot, only the value of a variable in the source code was changed in order to keep the actual compilation effort to a minimum and to emphasize the differences in the frameworks. All measurements were carried out five times and averaged. The error bars indicate the range of the measurement results.}
  \label{fig:rebuild_comp_comp}
\end{figure}

In local development, however, it is rare for a complete component or the full image to be built. It is much more common to trigger a rebuilt after changes to the source code are made. In such cases, build tools can use incremental build techniques to identify and build only those parts of the target whose source files have actually changed. An efficient implementation of such techniques can significantly accelerate the rebuild process, as can be seen in Fig.~\ref{fig:rebuild_comp_comp}. In all examples shown, only the value of a variable in the source code was changed. This keeps the actual compilation effort to a minimum and highlights how efficiently the respective framework handles the task. The results show that \ac{socks} achieves a higher efficiency than Yocto in all cases, with the largest difference of more than an order of magnitude in the case of the devicetree.

\begin{table}[t]
    \centering
    \caption{Disk space used by the successfully built example projects. The concept of downloading pre-built files instead of downloading the sources and compiling them locally has a significant impact on disk usage.}
    \label{tab:disk_usage}
    \begin{tabular}{cccc}
        \toprule
        Build Framework&Project Directory&Container Data (Docker)&Total\\
        \midrule
        \ac{socks}&4.2\,GB&6.2\,GB&10.4\,GB\\
        Yocto&78\,GB&-&78\,GB\\
        \bottomrule
    \end{tabular}
\end{table}

In addition to the pure build time of a project, the hard disk space required can also be a relevant metric. The reason for this is that a typical \ac{soc} image project can quickly grow to several tens of gigabytes in size, and if there are several projects on one system, this can occupy a significant part of the available storage space. Table~\ref{tab:disk_usage} shows the disk space usage of the example project implemented with \ac{socks} and with Yocto. The difference in size is also mainly due to the fact that Yocto compiles toolchains and file system elements from source code, while \ac{socks} downloads the corresponding components in precompiled format whenever possible. The size of these project directories is therefore only realized when the project is built. The size of the sources of the \ac{socks} image project itself, which must be included in version control, is only about 2.1\,MB. The remaining source data, such as the source files of the Linux kernel, are then automatically downloaded or generated at build time. Yocto uses a similar approach, which is why the corresponding source files are only 278\,MB. In comparison with \ac{socks}, this is significantly larger. The reason for this distinction is that the Yocto framework itself, including all required layers, is part of these project sources, either directly or as Git submodules. In contrast, \ac{socks}, with all its builders and container files, is installed as a Python package independently of the \ac{soc} image project. The source files for \ac{socks} are 30\,MB in size.

\section{Conclusion}

In this contribution, we introduced \ac{socks}, a lightweight and modular framework implemented in Python to simplify and accelerate the development workflow for complete \ac{soc} images. \ac{socks} was developed within the scientific community, which means that small teams of developers, potentially spread across several countries, are the target audience. Therefore, \ac{socks} is designed for distributed development and a workflow built around \ac{cicd}. The tool is open-source software and available at \cite{SocksGithub}.

By grouping the software components that form the bootable \ac{soc} image into the eponymous \ac{soc} blocks, a new abstraction layer is established. This simplifies the development process by dividing the effort into smaller, easier-to-handle modules. Clearly defined interfaces between the blocks and the reduced number of dependencies enable the \ac{soc} blocks to be built as independently as possible, which leads to a number of advantages. Exchanging modules of the same type is easily possible, and they can be shared between different projects, further reducing development effort. Furthermore, the modularization enables automated management of predefined build containers for every \ac{soc} block, which reduces the requirements on the software environment on the host system. Local development in containers also simplifies the creation of an associated \ac{cicd} pipeline for an \ac{soc} image project.

The presented measurement results show that \ac{socks} is in many scenarios---from processing small code modifications to full image builds---significantly faster than the established development tools. Depending on the host system used for building, a reduction in build time of up to 67 percent was observed. The use of a conventional Linux distribution, which does not have to be built from source code, contributes significantly to this result.

In future work, we plan to extend the \ac{socks} framework with additional \ac{soc} blocks and builders to support all hardware features of AMD's Zynq \ac{us+} and Versal families of devices. Specifically, the real-time processors embedded in both architectures and the AI engines available in a number of Versal devices.

\begin{acks}
This research acknowledges the support by the Doctoral School \emph{``Karlsruhe School of Elementary and Astroparticle Physics: Science and Technology''}. We thank Nicholas Tan Jerome for his helpful insights.
\end{acks}

\bibliographystyle{ACM-Reference-Format}
\bibliography{socks}

\end{document}